\documentclass{article}

\usepackage{amsmath}
\usepackage{times}
\usepackage{natbib}
\usepackage{color, amsmath, amsfonts, url, graphicx}
\usepackage[normalem]{ulem}
\usepackage{cancel}
\usepackage[utf8]{inputenc}
\usepackage[utf8]{inputenc}

\def\bone{{\bf 1}}
\def\ba{{\mbox{\boldmath$a$}}}

\def\bs{{\bf s}}

\def\bw{{\bf w}}
\def\bx{{\bf x}}
\def\by{{\bf y}}

\def\bA{{\bf A}}

\def\bO{{\bf O}}

\def\bS{{\bf S}}

\def\bW{{\bf W}}
\def\bX{{\bf X}}
\def\bY{{\bf Y}}

\def\thick#1{\hbox{\rlap{$#1$}\kern0.25pt\rlap{$#1$}\kern0.25pt$#1$}}
\def\balpha{\boldsymbol{\alpha}}

\def\bzeta{\boldsymbol{\zeta}}

\def\bphi{\boldsymbol{\phi}}

\def\bPsi{\boldsymbol{\Psi}}

\def\smbalpha{\boldsymbol{{\scriptstyle{\alpha}}}}

\def\ehat{{\widehat e}}

\def\Qhat{{\widehat Q}}
\def\Rhat{{\widehat R}}

\def\atilde{{\widetilde a}}

\def\gtilde{{\widetilde g}}

\def\otilde{{\widetilde o}}

\def\stilde{{\widetilde s}}

\def\ytilde{{\widetilde y}}

\def\Qtilde{{\widetilde Q}}

\def\gammahat{{\widehat\gamma}}
\def\deltahat{{\widehat\delta}}

\def\muhat{{\widehat\mu}}

\def\pihat{{\widehat\pi}}

\def\phihat{{\widehat\phi}}

\def\Deltahat{{\widehat\Delta}}

\def\Psihat{{\widehat\Psi}}

\def\phitilde{{\widetilde\phi}}

\def\bPsihat{{\widehat\bPsi}}
\def\smbalpha{\widehat{\smbalpha}}

\def\hbar{\bar{ h}}

\def\sbar{\bar{ s}}

\def\babar{\bar{ \ba}}

\def\bsbar{\bar{ \bs}}

\def\bwbar{\bar{ \bw}}

\def\bybar{\bar{ \by}}

\def\bAbar{\bar{ \bA}}

\def\bSbar{\bar{ \bS}}

\def\bWbar{\bar{ \bW}}

\def\bYbar{\bar{ \bY}}

\def\Isc{{\cal I}}

\def\Lsc{{\cal L}}

\def\Osc{{\cal O}}
\def\Psc{{\cal P}}
\def\Qsc{{\cal Q}}

\def\Usc{{\cal U}}

\def\Qschat{\widehat{{\cal Q}}}

\def\Psctilde{\widetilde{{\cal P}}}
\def\Qsctilde{\widetilde{{\cal Q}}}

\def\transpose{{\sf \scriptscriptstyle{T}}}

\def\half{\frac{1}{2}}

\def\nhalf{n^{\half}}

\def\nnhalf{n^{-\half}}

\def\E{\mbox{E}}

\def\var{\mbox{var}}

\def\sumin{\sum_{i=1}^n}

\def\trans{^{\transpose}}
\def\inv{^{-1}}

\def\var{\mbox{var}}

\def\mybox#1{\vskip1mm \begin{center}
        \hspace{.0\textwidth}\vbox{\hrule\hbox{\vrule\kern6pt
\parbox{.9\textwidth}{\kern6pt#1\vskip6pt}\kern6pt\vrule}\hrule}
        \end{center} \vskip-5mm}
\def\lboxit#1{\vbox{\hrule\hbox{\vrule\kern6pt
      \vbox{\kern6pt#1\vskip6pt}\kern6pt\vrule}\hrule}}

\def\thickboxit#1{\vbox{{\hrule height 1mm}\hbox{{\vrule width 1mm}\kern6pt
          \vbox{\kern6pt#1\kern6pt}\kern6pt{\vrule width 1mm}}
               {\hrule height 1mm}}}

\def\fat#1{\hbox{\rlap{$#1$}\kern0.25pt\rlap{$#1$}\kern0.25pt$#1$}}

\def\nnhalf{n^{-\half}}

\def\ba{{\mbox{\boldmath$a$}}}

\def\half{\frac{1}{2}}
\def\nhalf{n^{\half}}

\def\transpose{{\sf \scriptscriptstyle{T}}}

\def\trans{^{\transpose}}
\def\inv{^{-1}}
\def\Deltahat{{\widehat\Delta}}

\def\balpha{{\bf \alpha}}

     \def\E{\mbox{E}}

     \def\bW{\mathbf{W}}
  \def\bX{\mathbf{X}}
     
\def\bw{\mathbf{w}}

\def\sumin{\sum_{i=1}^n}

\def\Isc{{\cal I}}

\definecolor{green}{RGB}{000,150,100}
\definecolor{purple}{RGB}{150,000,180}

\def\bS{\mathbf{S}}
\def\bX{\mathbf{X}}
   \def\bs{\mathbf{s}}
    \def\bx{\mathbf{x}}

  \def\transpose{{\sf \scriptscriptstyle{T}}}

\def\supA{^{(A)}}
\def\supB{^{(B)}}
\def\supg{^{(g)}}

\def\Isc{\mathcal{I}}
\def\Usc{\mathcal{U}}

\def\supg{^{(g)}}

\def\E{\mathbb{E}}
\def\P{\mathbb{P}}
\def\supA{^{(1)}}
\def\supB{^{(0)}}
\def\Deltat{\Delta(t)}
\def\Deltastt{\Delta_\bS(t, t_0)}
\def\Rstt{R_\bS(t, t_0)}
\def\mubar{\bar{\mu}}
\def\Deltaht{\Deltahat(t)}
\def\Deltahstt{\Deltahat_\bS(t, t_0)}

 \def\trans{^{\transpose}}

\newcommand\independent{\protect\mathpalette{\protect\independenT}{\perp}}
\def\independenT#1#2{\mathrel{\rlap{$#1#2$}\mkern2mu{#1#2}}}

\usepackage{setspace}
\begin{document}
\newtheorem{assumption_2}{Assumption}
\newtheorem{prop}{Proposition}

\title{Robust Evaluation of Longitudinal Surrogate Markers with Censored Data}
\author{Denis Agniel$^{1*}$, Layla Parast$^2$\\
$^1$ RAND, Santa Monica, CA\\
$^2$ University of Texas, Austin, TX\\
$^*$ \url{dagniel@rand.org}}



\maketitle

\begin{abstract}
The development of statistical methods to evaluate surrogate markers is an active area of research. In many clinical settings, the surrogate marker is not simply a single measurement but is instead a longitudinal trajectory of measurements over time, e.g., fasting plasma glucose measured every 6 months for 3 years. In general, available methods developed for the single-surrogate setting cannot accommodate a longitudinal surrogate marker.  Furthermore, many of the methods have not been developed for use with primary outcomes that are time-to-event outcomes and/or subject to censoring. In this paper, we propose robust methods to evaluate a longitudinal surrogate marker in a censored time-to-event outcome setting. Specifically, we propose a method to define and estimate the proportion of the treatment effect on a censored primary outcome that is explained by the treatment effect on a longitudinal surrogate marker measured up to time $t_0$. We accommodate both potential censoring of the primary outcome and of the surrogate marker. A simulation study demonstrates good finite-sample performance of our proposed methods. We illustrate our procedures by examining repeated measures of fasting plasma glucose, a surrogate marker for diabetes diagnosis, using data from the Diabetes Prevention Program (DPP).
\end{abstract}


\section{Introduction} \label{intro}

In studies designed with long-term follow-up of study participants, identifying a surrogate marker that can be measured earlier to replace the primary outcome may allow for earlier decisions about new treatments. For example, early measurements of CD4 count and viral RNA are often examined as potential surrogate markers for progression to AIDS in studies of treatments among HIV patients \citep{o1996changes}. Many useful statistical methods have been developed to measure surrogacy, i.e., the strength of a surrogate, in terms of its ability to replace the primary outcome when evaluating a treatment effect \citep{prentice1989surrogate}. \citet{freedman1992statistical} proposed a straightforward approach to evaluate a single surrogate marker by estimating the proportion of the treatment effect on the primary outcome that is explained by the treatment effect on the surrogate marker via a regression model approach where the treatment effect is estimated both with and without adjusting for the surrogate marker. As an alternative, \citet{parast2016robust} and \citet{wang2020model} proposed to estimate this proportion using a robust kernel-based nonparametric approach. Several other measures have been proposed for the single-surrogate setting including the relative effect and adjusted association, average causal necessity, average causal sufficiency, and the causal effect predictiveness curve \citep{buyse1998criteria,gilbert2008evaluating,joffe2009related}. 

In many clinical settings, the surrogate marker is a longitudinal trajectory of measurements over time, e.g., fasting plasma glucose measured every 6 months for 3 years. In general, available methods developed for the single-surrogate setting cannot accommodate a longitudinal surrogate marker.  Furthermore, many of the methods have not been developed for use with primary outcomes that are time-to-event outcomes and/or subject to censoring. \citet{agniel2021evaluation}, for example, proposed flexible methods to estimate the proportion of the treatment effect explained by a longitudinal surrogate marker, but assumed a fully observed primary outcome with no censoring. When there is censoring, additional complexities are introduced as both the primary outcome and the surrogate marker may be censored. In this paper, we aim to fill this gap by addressing the setting where the surrogate is a longitudinal trajectory and the primary outcome is a censored time-to-event outcome.

\allowdisplaybreaks
Certainly, there exists a vast amount of literature on joint modeling of survival outcomes and longitudinal data \citep{rizopoulos2012joint,elashoff2016joint,crowther2013joint}. However, these methods are generally not within the framework of surrogate markers and do not attempt to define and estimate a metric that quantifies whether the longitudinal data captures the treatment effect on the primary outcome. Few methods have been proposed to evaluate a longitudinal surrogate marker in a setting with censored data.  With a censored outcome, \citet{renard2003validation} and  \citet{henderson2002identification} proposed measures of surrogacy within a meta-analytic setting relying on an unobserved latent zero-mean bivariate Gaussian process to describe the association between the longitudinal measurement and the event process. \citet{deslandes2007assessing} investigated both a multi-state model and a parametric joint model to estimate the proportion of treatment effect on the censored primary outcome that is explained by the longitudinal surrogate. \citet{zhou2023landmark} proposed a landmark survival model utilizing a varying coefficient model for the longitudinal measurement and quantified the variance explained by the measurements. Previous work has also explored evaluating a longitudinal surrogate within a Cox model framework, e.g., comparing two Cox models, one with versus without the surrogate or a joint model approach utilizing the Cox model for the censored outcome \citep{taylor2002surrogate,tsiatis1995modeling,dafni1998evaluating,zheng2022quantifying,liu2018exploring,zheng2022quantifying,le2022time}. While potentially useful, these available methods for a censored outcome involve parametric assumptions that are unlikely to hold in practice.

In this paper, we propose robust methods to evaluate a longitudinal surrogate marker in a censored time-to-event outcome setting. Specifically, we define the proportion of the treatment effect on a censored primary outcome that is explained by the treatment effect on a longitudinal surrogate marker measured up to time $t_0$ in terms of two treatment effects. We derive the efficient influence function for the treatment effect estimands, and we use these influence function to propose two estimators of the proportion of treatment effect explained: (1) a one-step plug-in estimator and (2) a targeted minimum loss-based (TML) estimator.  Our approach accommodates both potential censoring of the primary outcome and of the surrogate marker and imposes no parametric assumptions on the data-generating process. We investigate the finite-sample performance of our proposed methods using a simulation study and illustrate our procedures by examining repeated measures of fasting plasma glucose, a surrogate marker for diabetes diagnosis, using data from the Diabetes Prevention Program (DPP).

\section{Definitions and Influence Functions}
\subsection{Setting and Definitions} \label{definitions}
Let $G$ be a binary treatment indicator with $G=1$ for treatment and $G=0$ for control.  Let $T$ denote the time of the primary outcome, and let $\bS_{t_0}$  be the surrogate marker measurements up to some time $t_0$. Using potential outcomes notation, let $T^{(g)}$ and $\bS_{t_0}^{(g)}$ denote the time of the primary outcome and surrogate marker measurements under treatment $G = g$. In the absence of censoring, we only observe $(T, \bS_{t_0})=(T\supA, \bS_{t_0}\supA)$ or $(T\supB, \bS_{t_0}\supB)$ depending on whether $G=1$ or $0.$ Throughout, we assume that $T$ may take values in a discrete set $\{1, ..., t\}$ and $\bS_{t_0}^{(g)} = (S\supg_j)_{j=1,...,t_0}$ is defined (though not necessarily observed) at the first $t_0$ timepoints. Below, when there is no confusion, we will omit the subscript on $\bS$. Let $n_g = \sumin I\{G = g\}$ be the sample size in treatment group $G = g$.

We define the treatment effect on the primary outcome as the difference in survival rates at time $t\geq t_0$ between the two groups:
$$\Deltat = \P(T\supA > t) -  \P(T\supB > t).$$ Our aim is to quantify the proportion of this treatment effect on the primary outcome that can be explained by the treatment effect on the surrogate trajectory. If the proportion of treatment effect explained is high, this would likely indicate that this trajectory reflects a good surrogate, whereas a low proportion would indicate a poor surrogate. We build from the work of \citet{wang2002measure}, \citet{parast2017}, \citet{agniel2021evaluation}, and \citet{agniel2020doubly} and consider defining the residual treatment effect when the surrogate is a longitudinal marker and the primary outcome is a censored time-to-event outcome. The residual treatment effect is meant to capture the treatment effect that is ``leftover" after we account for the treatment effect on the longitudinal surrogate marker.

Our general strategy for defining this residual treatment effect is to first identify the overall treatment effect as a function of the surrogate:
    $\Deltat = \int \theta_1(\bs) dF_{\bS\supA}(\bs) - \int \theta_0(\bs) dF_{\bS\supB}(\bs)$
where $\theta_g(\bs)$ is the typical product-limit representation of the conditional survival function, which is integrated with respect to the treatment-specific distributions, $F_0, F_1$.
The residual treatment effect $\Deltastt$ can then be defined as 
    $\Deltastt = \int \left\{\theta_1(\bs) - \theta_0(\bs)\right\} dF^*_\bS(\bs)$,
a version of $\Deltat$ where the distribution of the surrogate is set to be equal between the treatment and control groups and integrated over a common distribution $F^*(\cdot)$. In the absence of censoring, $\Deltastt$ can be thought of as the following quantity
\begin{align*}
\int\left\{ \P\left( T \supA > t \mid \bS\supA = \bs\right) - \P\left( T \supB>t  \mid \bS\supB = \bs \right)\right\}dF^*_{\bS}(\bs) \label{rte1}
\end{align*}
 i.e., the expected difference in survival rates at time $t$, if the surrogate information up to $t_0$, is equivalent in the two groups. While there are many potential choices for $F^*_\bS(\cdot)$, we follow the approach in \citet{agniel2020doubly} and select $F^*_\bS(\cdot)$, the empirical distribution of $\bS$ including observations in both treatment groups.  This selection avoids choosing arbitrarily between either of the treatment-specific distributions $F_{\bS\supA}, F_{\bS\supB}$. It is also analogous to the choice of the reference distribution typically used in average treatment effect estimation, where confounders are integrated over their observed distribution, combining both treatment arms. 
 
If the treatment effect can be fully captured by the surrogate trajectory, then we would expect $\Delta_{\bS}(t,t_0)$, the residual treatment effect, to be zero. In contrast, if there is no treatment effect on the surrogate trajectory, we would expect  $\Delta_{\bS}(t,t_0)$ to be equal to $\Deltat$. The proportion of treatment effect explained by the longitudinal surrogate is then defined as:
 \begin{equation*} R_{\bS}(t,t_0)=\{\Deltat-\Delta_{\bS}(t,t_0)\}/\Deltat=1-\Delta_{\bS}(t,t_0)/\Deltat. \end{equation*}

Defining and estimating the residual treatment effect in this setting is complicated by the fact that some individuals may be censored or may experience the primary outcome before $t_0$. To fix ideas and build intuition, we first start in simpler settings before moving on to a setting of full generality. In all cases, estimation, asymptotics, and inference will be based on the influence functions for $\Deltat$ and $\Deltastt$. Therefore, we first derive the form of the influence functions in Sections \ref{sec:no-cens}, \ref{simplified-setting-w-censoring}, and \ref{sec:main-setting} before moving on to estimation and inference in Section \ref{sec:estimation-and-inf}. We first start with a simplified case with randomized treatment and no censoring (Section \ref{sec:no-cens}), then a simplified case with randomized treatment and censoring (Section \ref{simplified-setting-w-censoring}), and finally, the main setting with non-randomized treatment, many timepoints, and censoring in Section \ref{sec:main-setting}. 

\subsection{A simplified case with no censoring}\label{sec:no-cens}
We begin in a highly simplified setting where both the surrogate and the outcome are completely observed and the treatment is randomized. This means that there is no censoring of the outcome and that the outcome never occurs before $t_0$. In this case, we can write the relevant observed data as $O = (G, \bS, Y)$ where $Y = I\{T > t\}$ and $\bS = \bS_{t_0}$. This setting mirrors the setting in \citet{agniel2020doubly} and \citet{agniel2021evaluation} for (fully observed) longitudinal and multivariate surrogates. In this case, under Assumptions \ref{asmp:consistency}, \ref{asmp:pos}, and \ref{asmp:rand} given below, it is straightforward to show that $\Deltat = \E(Y | G = 1) - (Y | G = 0) = \int \mu_1(\bs)dF_{\bS | G = 1}(\bs) - \int \mu_0(\bs)dF_{S | G = 0}(\bs)$, and $\Deltastt = \int \left\{\mu_1(\bs) - \mu_0(\bs)\right\}dF_{\bS}(\bs) = \int \left\{\mu_1(\bs) - \mu_0(\bs)\right\}d\left\{\pi_1 F_{\bS | G = 1}(\bs) + \pi_0 F_{S | G = 0}(\bs)\right\}$
where $\pi_g = \P(G = g)$ and $\mu_g(\bs) = \E(Y | G = g, \bS = \bs)$. 

The assumptions required for identifying these causal effects from the observed data are:
\begin{assumption_2}[Consistency]\label{asmp:consistency}
$T\supg = T$ and $\bS\supg = \bS$ when $G = g$.
\end{assumption_2}
\begin{assumption_2}[Positivity]\label{asmp:pos}
$\P\{\pi_g(\bS) > \delta_1\} = 1$, where $\pi_g(\bs) = \P(G = g | \bS = \bs)$ for some $\delta_1 > 0$.
\end{assumption_2}
\begin{assumption_2}[Treatment randomization]\label{asmp:rand}
$T\supg, \bS\supg \independent G$
\end{assumption_2}

The influence function for $\Delta_\bS(t, t_0)$ is 
\begin{align*}
\phi\{O, \Delta_\bS(t, t_0), \bPsi\} =& \frac{GY - \{G - \pi_1(\bS)\}\mu_1(\bS)}{\pi_1(\bS)} - \frac{(1-G)Y - \{ 1-G - \pi_0(\bS)\}\mu_0(\bS)}{\pi_0(\bS)}
\end{align*}
for $\bPsi = \{\pi_0(\bs), \pi_1(\bs), \mu_0(\bs), \mu_1(\bs)\}$. This influence function can be used to develop doubly robust estimators for $\Deltastt$ \citep{bang2005doubly, Chernozhukov2017, agniel2020doubly} and -- along with similar results for $\Delta$ -- for $\Rstt$. See \citet{agniel2020doubly} for additional details.

\subsection{Simplified setting with censoring}\label{simplified-setting-w-censoring}
We now consider a more complex situation where both the surrogate and the outcome may experience censoring. Let $t = 2$ and $t_0 = 1$ and let the censoring time $C \in \{1, 2, \infty\}$, where $C = \infty$ if the person is not censored through time $t = 2$. Then the relevant observed data can be written as $O = (G, A_1, Y_1, S_1, A_2, Y_2)$ where $A_k = I\{C > k\}$ is an indicator that $Y_k$ is observed and $Y_k = I\{T > k\}, k = 1, 2$. We now require additional assumptions to identify $\Deltat$ and \begin{align*}
    \Deltastt = &\int \left\{\P(T\supA > t | S_1\supA = s, T\supA > 1)\P(T\supA > 1) - \right.\\
    &\qquad \left.\P(T\supB > t | S_1\supB = s, T\supB > 1)\P(T\supB > 1)\right\}dF_{S_1 | T > 1}(s).
\end{align*} 
The specification of $\Deltastt$ is a little less straightforward in this setting, as the surrogate is only observed for those who do not experience the event at time 1. The residual treatment effect standardizes the distribution of $S_1$ only among those where the surrogate is observed.

Define $\gamma_{g1} = \P(A_1 = 1 | G = g)$ and $\gamma_{g2}(s) = \P(A_2 = 1 | G = g, A_1 = 1, Y_1 = 1, S_1 = s)$ for the following assumptions. Here, we require:
\begin{assumption_2}[Positivity]\label{asmp:updated-pos}
$\P\{\pi_g(S_1) > \delta_1\} = 1$, where $\pi_g(s) = \P(G = g | A_1 = 1, Y_1 = 1, S_1 = s)$ for some $\delta_1 > 0$.
\end{assumption_2}
\begin{assumption_2}[Independent censoring]\label{asmp:indcens}
$T\supg \independent C | G, A_1 = 1, Y_1 = 1, S_1$ and $T\supg, S_1\supg \independent C | G$
\end{assumption_2}
\begin{assumption_2}[Censoring positivity]\label{asmp:censpos}
$\P\{\gamma_{g2}(S_1) > \delta_2\} = 1, \gamma_{g1} > \delta_2$ for $g = 0, 1$ and some $\delta_2 > 0$
\end{assumption_2}
Assumption \ref{asmp:updated-pos} is an updated version of Assumption \ref{asmp:pos} to account for potential censoring of the surrogate marker. It ensures that the surrogate distribution in the treatment groups has sufficient overlap. Assumption \ref{asmp:indcens} is the typical independent censoring assumption, stating that at each time point the censoring time and the failure time are independent conditional on all of the information available up to that point. Assumption \ref{asmp:censpos} ensures that the uncensored individuals are sufficiently similar to the censored individuals. 

Under Assumptions \ref{asmp:consistency}, \ref{asmp:rand}, \ref{asmp:indcens}, and \ref{asmp:censpos}, we show in Appendix \ref{app:ident-deltat} that there are multiple ways to identify $\Deltat$ from the observed data:
\begin{align*}
    \Deltat &= \mu_{11}\E\{\mu_{12}(S_1) | G = 1, A_1 = 1\} - \mu_{01}\E\{\mu_{02}(S_1) | G = 0, A_1 = 1\}\\
    &= \int \theta_1(s)dF_{S_1|G = 1, A_1 = 1, Y_1 = 1}(s) - \int \theta_0(s)dF_{S_1|G = 0, A_1 = 1, Y_1 = 1}(s)\\
    &= \int \theta_1(s)dF_{S_1\supA | Y_1\supA = 1}(s) - \int \theta_0(s)dF_{S_1\supB | Y_1\supB = 1}(s)\\
    &= \E\left\{n/n_1\frac{GA_1Y_1A_2Y_2}{\gamma_{11}\gamma_{12}(S_1)} - n/n_0\frac{(1-G)A_1Y_1A_2Y_2}{\gamma_{01}\gamma_{02}(S_1)} \right\}
\end{align*}
where $\mu_{g1} = \E(Y_1 | G = g, A_1 = 1), \mu_{g2}(s) = \E(Y_2 | G = g, A_1 = 1, Y_1 = 1, S_1 = s, A_2 = 1)$, and $\theta_g(s) = \mu_{g1}\mu_{g2}(s).$
The alternative forms of identifying $\Deltat$ above are suggestive of how one may identify $\Deltastt$. Whereas $\Deltat$ integrates $\theta_g(\bs)$ over the treatment-specific distribution of $S_1 | G = g, A_1 = 1, Y_1 = 1$, $\Deltastt$ must integrate over the common reference distribution. 

We show in Appendix \ref{app:ident-deltastt} that under Assumptions \ref{asmp:consistency}, \ref{asmp:rand}, \ref{asmp:updated-pos}, \ref{asmp:indcens}, and \ref{asmp:censpos},
\begin{align*}
    \Deltastt &= \int \left\{\theta_1(s) - \theta_0(s)\right\}dF_{S_1|A_1 = 1, Y_1 = 1}(s)\\
    &= \int \left\{\theta_1(s) - \theta_0(s)\right\}d\{\pi^*_1F_{S_1\supA|Y_1\supA = 1}(s) + (1-\pi^*_1)F_{S_1\supB|Y_1\supB = 1}(s)\}\\
    &= \E\left\{n/n_1\frac{GA_1Y_1A_2Y_2}{\gamma_{11}\gamma_{12}(S_1)} \times \frac{\pi^*_1}{\pi_1(S_1)} - n/n_0\frac{(1-G)A_1Y_1A_2Y_2}{\gamma_{01}\gamma_{02}(S_1)} \times \frac{1-\pi^*_1}{1-\pi_1(S_1)} \right\} 
\end{align*}
for $\pi_1^* = \P(G = 1 | A_1 = 1, Y_1 = 1)$.

We further show in Appendix \ref{if_delta_section} that the efficient influence function for $\Deltat$ is given by
\begin{align*}
    \phi\{O, \Deltat, \bPsi\} = &n/n_1G\left(
    \frac{A_1Y_1A_2}{\gamma_{11}\gamma_{12}(S_1)}\left\{Y_2 - \mu_{12}(S_1)\right\}  +
    \frac{A_1\mubar_{12}}{\gamma_{11}}\left\{Y_1 - \mu_{11}\right\} +
     \mu_{11}\mubar_{12}\right) - \\
    &\qquad n/n_0(1-G)\left(
    \frac{A_1Y_1A_2}{\gamma_{01}\gamma_{02}(S_1)}\left\{Y_2 - \mu_{02}(S_1)\right\} +
    \frac{A_1\mubar_{02}}{\gamma_{01}}\left\{Y_1 - \mu_{01}\right\} \right. \\ & \qquad \qquad \qquad \qquad \qquad \qquad\qquad\qquad\qquad\qquad\qquad \left. +
    \mu_{01}\mubar_{02}\right) - \Deltat
\end{align*}
where $\mubar_{g2} = \E(\mu_{g2}(S_1) | G = g, A_1 = 1, Y_1 = 1)$, and in Appendix \ref{if_deltas_section} we show that the efficient influence function for $\Deltastt$ is given by
\begin{align*}
    &\phi_S\{O, \Deltastt, \bPsi\} = \\
    &n/n_1G\left(
    \frac{A_1Y_1A_2\pi^*_1}{\gamma_{11}\gamma_{12}(S_1)\pi_1(S_1)}\left\{Y_2 - \mu_{12}(S_1)\right\}  +
    \frac{A_1\mu_{12}^*}{\gamma_{11}}\left\{Y_1 - \mu_{11}\right\} +
     \mu_{11}\mu_{12}^*\right) - \\
    &\qquad n/n_0(1-G)\left(
    \frac{A_1Y_1A_2\pi^*_0}{\gamma_{01}\gamma_{02}(S_1)\pi_0(S_1)}\left\{Y_2 - \mu_{02}(S_1)\right\} +
    \frac{A_1\mu_{02}^*}{\gamma_{01}}\left\{Y_1 - \mu_{01}\right\} +
    \mu_{01}\mu_{02}^*\right) + \\
    &\qquad A_1Y_1\left(\frac{n/n_1\pi_1^*}{\gamma_{11}}\left[\mu_{12}(S_1) - \mu_{12}^*\right] - \frac{n/n_0\pi_0^*}{\gamma_{01}}\left[\mu_{02}(S_1) - \mu_{02}^*\right] \right)- \Deltastt
\end{align*}
where $\mu_{g2}^* = \E(\mu_{g2}(S_1) | A_1 = 1, Y_1 = 1)$ and $\bPsi = (\gamma_{gk}, \mu_{gk}, \mubar_{gk}, \mu^*_{gk}, \pi_g, \pi^*_g)_{g = 0, 1; k = 1, 2}$ collects the nuisance functions. 

These influence functions can be used to develop efficient estimators for $\Delta$, $\Deltastt$, and $\Rstt$, but we first consider the most general setting before turning to estimation and inference.

\subsection{Non-randomized setting with censoring and many timepoints}\label{sec:main-setting}
We now consider the most general setting where treatment is not randomized and there are many potential timepoints ($t > 2$). The observed data vector can now  be written 
\begin{align*}
    O = (\bX, G, A_1, Y_1, S_1, A_2, Y_2, S_2, ..., A_{t_0}, Y_{t_0}, S_{t_0}, A_{t_0+1}, Y_{t_0+1}, A_{t_0+2}, ..., A_t, Y_t), 
\end{align*} 
where $A_k = I\{C > k\}, Y_k = I\{T > k\}$.
Define the history of the surrogate up through time $k$ as $\bSbar_k = (S_1, ..., S_k)$, with $\bSbar_k = \bSbar_{t_0}$ for $k \geq t_0$ and the history of the other variables similarly $\bAbar_k = (A_1, ..., A_k), \bYbar_k = (Y_1, ..., Y_k)$. The residual treatment effect now takes the form
\begin{align*}
    \Deltastt &= \int \left\{\prod_{k=1}^t\P(T\supA > k | T\supA > k - 1, \bSbar\supA_{k-1} = \bsbar_{k-1}) - \right.\\
    &\qquad\qquad \left.\prod_{k=1}^t\P(T\supB > t | T\supB > k - 1, \bSbar\supB_{k-1} = \bsbar_{k-1})\right\}d\left\{\prod_{k=1}^{t_0} F_{S_k | T > k}(s_k)\right\}.
\end{align*}
We now require the additional assumptions:
\begin{assumption_2}[Positivity]\label{asmp:pos2}
$\P\{\delta_1 < \pi_{k}(\bX, \bSbar_k) < 1-\delta_1\} = 1$, $\P\{\delta_1 < e(\bX) < 1-\delta_1\} = 1$, for $e(\bx) = \P(G = 1 | \bX = \bx), \pi_{k}(\bx, \bsbar_k) = \P(G = 1 | \bX = \bx, \bAbar_k = \bYbar_k = 1, \bSbar_k = \bsbar_k), k = 1, ..., t-1$ and some $\delta_1 > 0$.
\end{assumption_2}
\begin{assumption_2}[Treatment ignorability]\label{asmp:ign}
$T\supg, \bS\supg \independent G | \bX, g = 0, 1$
\end{assumption_2}
\begin{assumption_2}[Independent censoring]\label{asmp:indcens2}
$T\supg \independent C | \bSbar_k, \bX, G, k = 0, ..., t-1, g = 0, 1$
\end{assumption_2}
\begin{assumption_2}[Censoring positivity]\label{asmp:censpos2}
$\P\{\gamma_{gk}(\bX, \bSbar_{k-1}) > \delta_2\} = 1$ for $\gamma_{gk}(\bx, \bsbar_{k-1}) = \P(A_k = 1 | \bX = \bx, G = g, \bAbar_{k-1} = \bYbar_{k-1} = 1, \bSbar_{k-1} = \bsbar_{k-1}), g = 0, 1, k = 1, ...., t$ and some $\delta_2 > 0$
\end{assumption_2}

These assumptions directly generalize assumptions from previous sections to this most general setting. For example, Assumption \ref{asmp:pos2} generalizes Assumption \ref{asmp:pos}, Assumption \ref{asmp:ign} generalizes Assumption \ref{asmp:rand} to allow for confounding by pre-treatment covariates, Assumption \ref{asmp:indcens2} generalizes Assumption \ref{asmp:indcens}, and Assumption \ref{asmp:censpos2} generalizes Assumption \ref{asmp:censpos}.

Under Assumptions \ref{asmp:consistency},  \ref{asmp:ign}, \ref{asmp:indcens2}, \ref{asmp:censpos2}, we show in Appendix \ref{app:main-setting-id-deltat} that
\begin{align*}
    \Deltat &= \Qsc_{10} - \Qsc_{00} \\
    &= \E_\bX\left\{\mu_{11}(\bX)\Qsc_{11}(\bX) - \mu_{01}(\bX)\Qsc_{01}(\bX)\right\}\\
    &= \E\left\{ 
                    \frac{G}{e(\bX)} \prod_{k=1}^t \frac{A_{k}Y_{k}}{\gamma_{1k}(\bX, \bSbar_{k-1})} - \frac{1-G}{1-e(\bX)} \prod_{k=1}^t \frac{A_{k}Y_{k}}{\gamma_{0k}(\bX, \bSbar_{k-1})}
                  \right\}
\end{align*}
    where for $k = 1, ..., t_0$,
\begin{align*}
    \Qsc_{gk}(\bx, \bsbar_{k-1}) = \E_{S_k}\{\mu_{gk+1}(\bx, \bSbar_{k})\Qsc_{gk+1}(\bx, \bSbar_{k}) | \bX = \bx, G = g, \bAbar_{k-1} = \bYbar_{k-1} = 1, \bSbar_{k-1} = \bsbar_{k-1}\},
\end{align*}
$\mu_{gk}(\bx, \bsbar_{k-1}) = \E\{Y_k | \bX = \bx, G = g, \bAbar_{k} = \bYbar_{k-1} = 1, \bSbar_{k-1} = \bsbar_{k-1}\}$, and for $k = t_0+1, ... t-1, \Qsc_{gk}(\bx, \bsbar_{t_0}) = \prod_{j=k+1}^t\mu_{gk+1}(\bx, \bsbar_{t_0})$, and $\Qsc_{gt}(\bx, \bsbar_{t-1}) = 1$.
    Furthermore, we show in Appendix \ref{app:main-setting-id-deltastt} that under Assumptions \ref{asmp:consistency}, \ref{asmp:pos2}, \ref{asmp:ign}, \ref{asmp:indcens2}, \ref{asmp:censpos2}
\begin{align*}
    \Deltastt &= \Qsc^*_{10} - \Qsc^*_{00} \\
    &= \E\left\{ 
                    \frac{G}{e(\bX)} \frac{A_{1}Y_{1}}{\gamma_{11}(\bX)}\prod_{k=2}^{t_0+1} \frac{A_{k}Y_{k}\pi^*_{k-1}(\bX, \bSbar_{k-2})}{\gamma_{1k}(\bX, \bSbar_{k-1})\pi_{k-1}(\bX, \bSbar_{k-1})}\prod_{k=t_0+2}^t \frac{A_{k}Y_{k}}{\gamma_{1k}(\bX, \bSbar_{t_0})} -\right.\\
                    &\qquad\left.\frac{1-G}{1-e(\bX)}\frac{A_{1}Y_{1}}{\gamma_{01}(\bX)} \prod_{k=2}^{t_0+1} \frac{A_{k}Y_{k}\{1-\pi^*_{k-1}(\bX, \bSbar_{k-2})\}}{\gamma_{0k}(\bX, \bSbar_{k-1})\{1-\pi_{k-1}(\bX, \bSbar_{k-1})\}}
                    \prod_{k=t_0+2}^t \frac{A_{k}Y_{k}}{\gamma_{0k}(\bX, \bSbar_{t_0})}
                  \right\}
\end{align*}
where 
\begin{align*}
    \Qsc^*_{g1}(\bx) &= \E_{S_1}\{\mu_{g2}(\bx, S_1)\Qsc^*_{g2}(\bx, S_1)|\bX = \bx, A_1 = Y_1 = 1\}\\
    \Qsc^*_{gk}(\bx, \bsbar_{k-1}) &= \E_{S_k}\left\{\mu_{gk+1}(\bx, \bSbar_{k})\Qsc^*_{gk+1}(\bx, \bSbar_{k}) | \bX = \bx, \bAbar_{k-1} = \bYbar_{k-1} = 1, \bSbar_{k-1} = \bsbar_{k-1} \right\}, \\
    &~~~~~~~~~~~~~~~~~~~~~~~~~~~~~~~~~~~~~~~~~~~~~~~~~~k = 2, ..., t_0+1\\
    \Qsc^*_{gk}(\bx, \bsbar_{t_0}) &= \prod_{j=k+1}^t\mu_{gk+1}(\bx, \bsbar_{t_0}) = \Qsc_{gk}(\bx, \bsbar_{t_0}), k = t_0+2, ..., t - 1\\
    \Qsc^*_{gt}(\bx, \bsbar_{t_0}) &= 1 = \Qsc_{gt}(\bx, \bsbar_{t_0}).
\end{align*}
and $\pi^*_{gk}(\bx, \sbar_{k-1}) = \P(G = g | \bX = \bx, \bAbar_{k} = \bYbar_{k} = 1, \bSbar_{k-1} = \sbar_{k-1})$.

Appendix \ref{if_delta_second} also shows that the influence function for $\Deltat$ may be written as: 
\begin{align*}
    &\phi\{O, \Deltat, \bPsi\} =\\
    &\frac{G}{e}
                             \left(
                                \sum_{j=1}^t \prod_{k=1}^{j-1} \frac{A_{k}Y_{k}}{\gamma_{1k}}
                                \left[
                                    \frac{A_j}{\gamma_{1j}}\{Y_j\mu_{1j+1}\Qsc_{1j+1} - \mu_{1j}\Qsc_{1j}\}
                                \right]
                                \right)  -\\
                               &\qquad  \frac{1-G}{1-e}
                             \left(
                                \sum_{j=1}^t \prod_{k=1}^{j-1} \frac{A_{k}Y_{k}}{\gamma_{0k}}
                                \left[
                                    \frac{A_j}{\gamma_{0j}}\{Y_j\mu_{0j+1}\Qsc_{0j+1} - \mu_{0j}\Qsc_{0j}\}
                                \right]
                                \right)\\
                                &\qquad\qquad  + \mu_{11}\Qsc_{11} - \mu_{01}\Qsc_{01} - \Deltat
\end{align*}
where for notational compactness we suppress the arguments of the functions, but we refer the reader to Table \ref{tab:defs} for all definitions and dependencies. The influence function for $\Deltastt$ can similarly be written as 
\begin{align*}
    \phi_S\{O, \Deltastt, \bPsi\} =
    & \frac{G}{e}
                             \left(
                                \sum_{j=1}^{t} \prod_{k=1}^{j-1} \frac{A_{k}Y_{k}\pi^*_{k-1}}{\gamma_{1k}\pi_{k-1}}
                                \left[
                                    \frac{\pi^*_{j-1}A_j}{\pi_{j-1}\gamma_{1j}}
                                \right]\Qsc_{1j}^*\{Y_j - \mu_{1j}\}\right)  -\\
                                &\qquad \frac{1-G}{1-e}
                             \left(
                                \sum_{j=1}^{t} \prod_{k=1}^{j-1} \frac{A_{k}Y_{k}\{1-\pi^*_{k-1}\}}{\gamma_{0k}\{1-\pi_{k-1}\}}
                                \left[
                                    \frac{\{1-\pi^*_{j-1}\}A_j}{\{1-\pi_{j-1}\}\gamma_{0j}}
                                \right]\Qsc^*_{0j}\{Y_j - \mu_{0j}\}
                                \right) +\\
                                &\qquad\qquad e\inv \sum_{j=1}^{t_0}\prod_{k=1}^{j}\frac{\pi^*_{k-1}A_kY_k}{\pi_{k-1}\gamma_{1k}}\pi^*_{j}\left\{\mu_{1j+1}\Qsc^*_{1j+1} - \Qsc^*_{1j}\right\} - \\
                                &\qquad\qquad (1-e)\inv \sum_{j=1}^{t_0}\prod_{k=1}^{j}\frac{(1-\pi^*_{k-1})A_kY_k}{(1-\pi_{k-1})\gamma_{0k}}(1-\pi^*_{j})\left\{\mu_{0j+1}\Qsc^*_{0j+1} - \Qsc^*_{0j}\right\} + \\
                                &\qquad\qquad \mu_{11}\Qsc^*_{11} - \mu_{01}\Qsc^*_{01} - \Deltastt
\end{align*}
where 
$\pi^*_j = \pi_j = 1$ for $j > t_0$, and again $\bPsi$ collects all nuisance functions $\bPsi = (e, \pi_k, \pi^*_k, \gamma_{gk}, \mu_{gk}, \Qsc_{gk}, \Qsc^*_{gk})_{g = 0, 1;, k = 1, ..., t}$. See Appendix \ref{if_deltas_second} for details.

\begin{center}
\begin{table}[]     \caption{Definition of key quantities.}
    \label{tab:defs}

    \centering
\hspace*{-0.3in}    \begin{tabular}{|l| l|} 
 \hline
 \textbf{Quantity} & \textbf{Definition} \\ 
 \hline
 $\bX$ & Pre-treatment covariates\\
 \hline
 $G$ & Treatment arm indicator\\
 \hline
 $A_k$ & Indicator that observation is not censored at time $k$\\ 
 \hline
 $Y_k$ & Survival through time $k$ \\
 \hline
 $S_k$ & Surrogate at time $k$ \\
 \hline
 $\bS$ & Surrogate marker measurements up to $t_0$\\
 \hline
 $\Delta(t)$ & Treatment effect, $\P(T\supA > t) -  \P(T\supB > t)$\\ 
 \hline
 $ \Deltastt$ & Residual treatment effect \\
 \hline
 $R_{\bS}(t,t_0)$ &  Proportion of treatment effect explained by $\bS$\\ 
 \hline
 $e(\bx)$ & $\P(G = 1 | \bX = \bx)$ \\
 \hline
 $\pi_{k}(\bx, \bsbar_k)$ & $\P(G = 1 | \bX = \bx, \bAbar_k = \bYbar_k = 1, \bSbar_k = \bsbar_k)$  \\
 \hline
 $\pi^*_k(\bx, \bsbar_{k-1})$ & $\P(G = 1 | \bX = \bx, \bAbar_k = \bYbar_k = 1, \bSbar_{k-1} = \bsbar_{k-1})$  \\
 \hline
 $\gamma_{gk}(\bx, \bsbar_{k-1})$ & $\P(A_k = 1 | \bX = \bx, G = g, \bAbar_{k-1} = \bYbar_{k-1} = 1, \bSbar_{k-1} = \bsbar_{k-1})$  \\
 \hline
 $\mu_{gk}(\bx, \bsbar_{k-1})$ & $\E\{Y_k | \bX = \bx, G = g, \bAbar_{k} = \bYbar_{k-1} = 1, \bSbar_{k-1} = \bsbar_{k-1}\}$ \\
 \hline
 $\Qsc_{gk}(\bx, \bsbar_{k-1})$ & $\E_{S_k}\left\{\mu_{gk+1}(\bx, \bSbar_{k})\Qsc_{gk+1}(\bx, \bSbar_{k}) | \bX = \bx, G = g, \bAbar_{k-1} = \bYbar_{k-1} = 1, \bSbar_{k-1} = \bsbar_{k-1} \right\}$\\
 \hline
 $\Qsc^*_{gk}(\bx, \bsbar_{k-1}),$ & $\E_{S_k}\left\{\mu_{gk+1}(\bx, \bSbar_{k})\Qsc^*_{gk+1}(\bx, \bSbar_{k}) | \bX = \bx, \bAbar_{k-1} = \bYbar_{k-1} = 1, \bSbar_{k-1} = \bsbar_{k-1} \right\}$\\
 $~~k \leq t_0+1$ & \\
 \hline
 $\Qsc^*_{gk}(\bx, \bsbar_{k-1}),$ & $\Qsc_{gk}(\bx, \bsbar_{k-1})$\\
 $~~k>t_0+1$ & \\[1ex] 
 \hline
 $\bPsi$ & $(e, \pi_k, \pi^*_k, \gamma_{gk}, \mu_{gk}, \Qsc_{gk}, \Qsc^*_{gk})_{g = 0, 1;, k = 1, ..., t}$\\
 \hline
\end{tabular}
\end{table}
\end{center}

\section{Robust Estimation and Inference}\label{sec:estimation-and-inf}
\subsection{Estimation}
We propose two estimators of $\Deltat$ and $\Deltastt$ based on the influence functions: (1) a one-step plug-in and (2) a targeted minimum loss-based (TML) estimator \citep{van2012targeted,zheng2017longitudinal,diaz2021nonparametric}. The plug-in estimator is straightforward to implement: once estimates of all nuisance functions have been estimated -- call these estimates $\bPsihat$ -- the estimator simply plugs these nuisance functions into the influence function and solves for $\Deltastt$ or $\Deltat$. These nuisance functions include the propensity score $e$, the ``propensity" scores $\pi_k^*(\bx, \bsbar_{k-1}), \pi_k(\bx, \bsbar_{k})$, the non-censoring probabilities $\gamma_{gk}(\bx, \bsbar_{k-1})$, the hazards $\mu_{gk}(\bx, \bsbar_{k-1})$, and the mean functions $\Qsc_{gk}(\bx, \bsbar_{k-1}), \Qsc^*_{gk}(\bx, \bsbar_{k-1})$. Estimation of these terms is required to evaluate the influence function and estimate the parameters of interest. 

In some settings, parametric models could in principle be used to estimate these quantities. However, to make these estimators as generally applicable as possible, we propose to use machine learning estimators along with cross-fitting \citep{diaz2021nonparametric,agniel2021evaluation,Chernozhukov2017, kennedy2022semiparametric} to separate the estimation of nuisance functions from their evaluation in the influence function.

Without loss of generality, suppose we split the dataset into two parts. Let $\Osc = (\bO_i)_{i = 1, 2, ..., n}$ be the observed data for analysis. Take $\Isc_0, \Isc_1$ to be a partition of the indices $\{1, 2, ..., n\}$, and call $\Osc_0 = (\bO_i)_{i \in \Isc_0}, \Osc_1 = (\bO_i)_{i \in \Isc_1}$. Finally, let $\bPsihat_\ell$ be an estimate of $\bPsi$ using data in $\Osc_\ell$. We propose to estimate $\Rhat_S(t, t_0)$ using a one-step plug-in estimator, $\Rhat_S(t, t_0) = 1 - \Deltahstt/\Deltaht$ where
\begin{align*}
    \Deltahstt = n\inv \sum_{\ell = 0}^1 \sum_{i \in \Isc_{\ell}} \phitilde_S(O_i, \bPsihat_{1 - \ell}), \quad \Deltaht = n\inv \sum_{\ell = 1}^2 \sum_{i \in \Isc_{\ell}} \phitilde(O_i, \bPsihat_{1 - \ell}),
\end{align*}
$\phitilde(O, \bPsi) = \phi(O, \bPsi) + \Deltat$ is an uncentered version of the influence function, and similarly $\phitilde_S(O, \bPsi) = \phi_S(O, \bPsi) + \Deltastt$. Extensions to more than two splits are straightforward.

Despite its relatively direct implementation, the one-step estimator may have drawbacks in some settings. A major drawback is that the one-step estimator could yield estimates of $\P(T\supg > t)$ that do not fall in $[0, 1]$ and in general may be subject to instability if estimates of $\gamma_{gk}$ or $\pi_k$ get too near to 0 or 1. To address these potential shortcomings, we additionally propose a TML estimator \citep{van2006targeted} which is computed using a similar approach as in \citet{diaz2021nonparametric} and \citet{zheng2017longitudinal}, using iterative logistic tilting models. For both $\Delta(t)$ and $\Deltastt$, estimation involves cross-fitting followed by a logistic tilting model that ensures the resulting estimator solves the influence function to obtained targeted versions of the quantities of interest. Full details describing the estimation algorithm are given in Appendix \ref{app:tmle}.

\subsection{Asymptotic distribution and inference}

The asymptotic behavior of the two sets of estimators for $\Deltat, \Deltastt$, and $R_S(t, t_0)$ depend on the good behavior of the estimation of the nuisance functions. If the estimation of the nuisance functions is fast enough, then the variability due to estimating them is asymptotically negligible, and estimates of  $\Deltat, \Deltastt$, and $R_S(t, t_0)$ obtain parametric rates.

First, we show in Appendix \ref{app:asymptotics} that $\Deltat$ and $\Deltastt$ have ``doubly robust" properties in the sense that one need not correctly specify all nuisance parameters to obtain an influence function with mean 0. Since estimation is based on solving this influence function, this property ensures that estimation is correctly targeted. 

Let $\phihat = \phi\{\bO, \Deltat, \Psihat\}, \phihat_\bS = \phihat_\bS(\bO, \Deltastt, \Psihat)$ be the estimated influence function for the overall and residual treatment effects, respectively, for some nuisance function estimates $\Psihat$. We have the following result:
\begin{prop}\label{prop_dr}
    Consider the following conditions:
    \begin{enumerate}
        \item[(a)] $\muhat_{gk}(\bx, \bsbar_{k-1}) = \mu_{gk}(\bx, \bsbar_{k-1})$.\\
        \item[(b)] $\Qschat_{gk}(\bx, \bsbar_{k-1}) = \Qsc_{gk}(\bx, \bsbar_{k-1}).$
        \item[(c)] $\Qschat^*_{gk}(\bx, \bsbar_{k-1}) = \Qsctilde_{gk}(\bx, \bsbar_{k-1})$
        for $$\Qsctilde_{gk}(\bx, \bsbar_{k-1}) = \E\left\{\Qschat^*_{gk+1}(\bx, \bsbar_k)\muhat_{gk+1}(\bx, \bsbar_k) | \bX = \bx, \bAbar_{k} = \bYbar_k = 1, \bSbar_{k-1} = \bsbar_{k-1}\right\}$$.
        \item[(d)] $\ehat(\bx) = e(\bx).$
        \item[(e)] $\gammahat_{gk}(\bx, \bsbar_{k-1}) = \gamma_{gk}(\bx, \bsbar_{k-1}).$ 
        \item[(f)] $\pihat_k(\bx, \bsbar_k) = \pi_k(\bx, \bsbar_k).$\\
        \item[(g)] $\pihat^*_k(\bx, \bsbar_{k-1}) = \pi^*_k(\bx, \bsbar_{k-1})$.
    \end{enumerate}
    If either (a) and (b) hold or (d) and (e) hold, we have that $\E(\phihat) = 0.$ And if either (a) and (c) hold or (d), (e), (f), and (g) hold, we have that $\E(\phihat_S) = 0.$
\end{prop}
Proof is given in Appendix \ref{app:asymptotics}. This establishes the double robustness property of estimators based on these influence functions, similar to what has been found previously for related mediation estimators \citep{van2012targeted}. It states that, if \textit{either} the models for the outcome are correctly specified \textit{or} the models for the weighting (the propensity score, censoring model, and the models that reweight the surrogate distribution), the influence function is centered on the true parameters. Thus, estimators based on the influence function can be expected under regularity conditions to have nice properties. One could alternatively propose``sequentially double robust" versions of these estimators \citep{luedtke2017sequential}, a less restrictive version of double robustness, but we leave this for future work.

In general, we do not expect to specify statistical models that are entirely correct for any of the nuisance functions. However, a result of the double robustness established by Proposition \ref{prop_dr} is that the errors in the outcome functions are in some sense offset by the errors in the weighting functions. This means that, if sample-splitting is used, estimation of the nuisance functions may be much slower than the parametric $\nnhalf$ rate and that therefore machine learning methods may be used to estimate all nuisance functions. 

Specifically, we have the following proposition. 
\begin{prop}
~
\begin{enumerate}
    \item[(i)] Under Assumptions \ref{asmp:consistency},  \ref{asmp:ign}, \ref{asmp:indcens2}, \ref{asmp:censpos2}, if estimators of the nuisance functions obtain the following convergence rates
\begin{align*}
    &\E\left(\sum_{g = 0}^1\sum_{k=1}^t\{\Qschat_{gk}(\bX, \bSbar_{k-1}) - \Qsc_{gk}(\bX, \bSbar_{k-1})\}^2 + \{\muhat_{gk}(\bX, \bSbar_{k-1}) - \mu_{gk}(\bX, \bSbar_{k-1})\}^2\right) \times \\
    &\qquad \E\left[\left\{\ehat(\bX) - e(\bX)\right\}^2 + \sum_{g = 0}^1\sum_{k=1}^t \left\{\gammahat_{gk}(\bX, \bSbar_{k-1}) - \gamma_{gk}(\bX, \bSbar_{k-1})\right\}^2\right] = o_p(\nnhalf),
\end{align*}
then we have 
\begin{align*}
    \nhalf\{\Deltaht - \Deltat\} \longrightarrow N(0, \sigma^2_\Delta)
\end{align*}
where $\sigma^2_\Delta = \var[\phi\{O, \Deltat, \theta\}].$
\item[(ii)] Under Assumptions \ref{asmp:consistency}, \ref{asmp:pos2}, \ref{asmp:ign}, \ref{asmp:indcens2}, and \ref{asmp:censpos2}, if estimators of the nuisance functions obtain the following convergence rates
\begin{align}
\begin{aligned}\label{convergence-rates}
    &\E\left(\sum_{g = 0}^1\sum_{k=1}^t\{\Qschat^*_{gk}(\bX, \bSbar_{k-1}) - \Qsctilde_{gk}(\bX, \bSbar_{k-1})\}^2 + \{\muhat_{gk}(\bX, \bSbar_{k-1}) - \mu_{gk}(\bX, \bSbar_{k-1})\}^2\right) \times \\
    &\qquad \E\left[\left\{\ehat(\bX) - e(\bX)\right\}^2 + \sum_{g = 0}^1\sum_{k=1}^t \left\{\gammahat_{gk}(\bX, \bSbar_{k-1}) - \gamma_{gk}(\bX, \bSbar_{k-1})\right\}^2 + \sum_{k=1}^{t_0}\left\{\pihat_{k}(\bX, \bSbar_{k}) - \pi_{k}(\bX, \bSbar_{k})\right\}^2 + \right.\\
    &\qquad \left.\sum_{k=1}^{t_0}\left\{\pihat^*_{k}(\bX, \bSbar_{k-1}) - \pi^*_{k}(\bX, \bSbar_{k-1})\right\}^2\right] = o_p(\nnhalf),
    \end{aligned}
\end{align}
then we obtain 
\begin{align*}
    \nhalf\{\Deltahstt - \Deltastt\} \longrightarrow N(0, \sigma^2_{\Delta_S})
\end{align*}
where $\sigma^2_{\Delta_S} = \var[\phi_S\{O, \Deltastt, \theta\}]$.
\end{enumerate}
\end{prop}
These results are shown in Appendix \ref{app:asymptotics} and state that the one-step and TML estimators have good behavior in most typical settings. Because of cross-fitting, the machine learning estimators of the nuisance functions are only required to converge at the rate $n^{-1/4}$, a rate that most off-the-shelf algorithms obtain. The asymptotic distribution of the estimators is Normal, with variance proportional to the variance of the influence function, which may be estimated by the variance of the empirical influence function.

Finally, these results imply the asymptotic distribution of the PTE to be:
\begin{align*}
    \nhalf\{\Rhat_S(t, t_0) - R_S(t, t_0)\} \longrightarrow N(0, \sigma^2_R)
\end{align*}
with 
\begin{align}\label{sigma}
\begin{aligned}
\sigma^2_R &= \Delta(t)^{-2}\E\left[\phi\{\bO, \Delta(t), \bPsi\}^2\right] + \Delta_{\bS}(t, t_0)^2\Delta(t)^{-4}\E\left[\phi_S\{\bO, \Delta_\bS(t, t_0), \bPsi\}^2\right] - \\
&\qquad 2\Delta_{\bS}(t, t_0)\Delta(t)^{-3}\E\left[\phi\{\bO, \Delta(t), \bPsi\}\phi_S\{\bO, \Delta_\bS(t, t_0), \bPsi\}\right].    
\end{aligned}
\end{align}
This variance may also be estimated by plugging in $\Deltaht, \Deltahstt$ and evaluating the empirical versions of the expectations in \eqref{sigma}.

\section{Simulation Study}

The aim of this simulation study was to examine the finite sample performance of our proposed methods with respect to bias and confidence interval coverage. All of our proposed methods are implemented in the R package \texttt{survivalsurrogate} available at \url{https://github.com/denisagniel/survivalsurrogate}. 

While potential comparison methods are described in Section \ref{intro}, they are generally infeasible to implement or are not comparable to our approach. Specifically, the vast amount of prior work on general joint modeling of survival outcomes and longitudinal data \citep{rizopoulos2012joint,elashoff2016joint,crowther2013joint} do not define or estimate a metric like the PTE for a surrogate marker, as discussed here. Meta-analytic methods such as \citet{renard2003validation} and  \citet{henderson2002identification} are not applicable to our single-study setting. The Cox model-based methods that do focus on examining surrogacy have two limitations: a) they do not examine the treatment effect $\Delta(t)$, as we do here, but instead focus on the hazard ratio for treatment, and b) do not offer reproducible code to facilitate comparisons \citep{taylor2002surrogate,tsiatis1995modeling,dafni1998evaluating}. While not focused on the PTE quantity, a more recent paper, \citet{zheng2022quantifying}, does propose a useful approach to examine the treatment effect and indirect effect with respect to a longitudinal marker via causal mediation analysis; unfortunately, their provided code is infeasible to implement in \texttt{R} given the use of a separate SAS code needed to stably handle the optimization with integration over
the random effect distribution when estimating the parameter from their new joint model -- which is different from the model that can be fitted with the \texttt{JM} package.  For these reasons, we do not compare our estimators to available approaches in this simulation study. However, in our application to real data in Section \ref{sec:example}, we do offer a comparison to a general joint model approach that focuses on an alternative quantity, the hazard within a Cox model framework via the \texttt{JMbayes} package in \texttt{R}; details are provided in Appendix \ref{app:sims}.

We examined performance in three simulation settings. For all settings,  $n=1000$, $t=6$ and $t_0=5$. Censoring was generated from an exponential distribution. In Setting 1, the surrogate up to $t_0$ was almost useless such that $R_{\bS}(t,t_0) = 0.028$. In Setting 2, the surrogate up to $t_0$ was nearly perfect such that $R_{\bS}(t,t_0) = 0.966$. In Setting 3, the surrogate up to $t_0$ was between these two extremes such that the true $R_{\bS}(t,t_0) = 0.603$. Data generated (via the \texttt{simcausal} package \citep{simcausal}) in all settings was non-randomized and had the surrogate measured at 5 time points: $(1,2,3,4,5)$. Simulation setup details are provided in Appendix \ref{app:sims}. Simulation results are shown in Table \ref{sim_results}. In Settings 1 and 3, the bias for $\Delta(t)$, $\Deltastt$, and $R_{\bS}(t,t_0)$ for both estimators is small, though the bias when estimating $R_{\bS}(t,t_0)$ is smaller for the TML estimator e.g. 0.015 (plug-in) vs. -0.007 (TML). In Setting 2, where $R_{\bS}(t,t_0) = 0.966$, results show higher bias for the TML estimator compared to the plug-in estimator for estimating $\Deltastt$, and $R_{\bS}(t,t_0)$.  Coverage levels are close to the nominal level of 0.95 with some slight deviations between 0.91-0.98 in some cases. For example, the confidence intervals for the TML estimator of $R_{\bS}(t,t_0)$ show slight overcoverage (0.975) in Setting 1 and slight undercoverage (0.919) in Setting 2. Overall, these results demonstrate reasonable performance of these proposed estimators in terms of minimal bias and coverage.

\begin{table}[hptb]
\caption{Simulation Results in Settings 1-3 for the proposed plug-in estimator and TML estimator; Bias is the average of the difference between the estimate and the truth; CI coverage is the proportion of iterations where the confidence interval contained the truth \label{sim_results}}
\begin{center}
\begin{tabular}{|l|c|c|c|c|} \hline
&\multicolumn{4}{c|}{Setting 1}\\ \hline
&\multicolumn{2}{c|}{Plug-in Estimator}&\multicolumn{2}{c|}{TML Estimator}\\ \hline
&\multicolumn{1}{c|}{Bias}&\multicolumn{1}{c|}{CI Coverage}&\multicolumn{1}{c|}{Bias}&\multicolumn{1}{c|}{CI Coverage}\\ \hline
$\Delta(t)$     &0.004            &0.965            & 0.001           &0.971            \\ 
$\Deltastt$     &0.000            &0.960            & 0.003           &0.972            \\ 
$R_{\bS}(t,t_0)$&0.015            &0.943            &-0.007           &0.975            \\ 
  \hline
&\multicolumn{4}{c|}{Setting 2}\\ \hline
&\multicolumn{2}{c|}{Plug-in Estimator}&\multicolumn{2}{c|}{TML Estimator}\\ \hline
&\multicolumn{1}{c|}{Bias}&\multicolumn{1}{c|}{CI Coverage}&\multicolumn{1}{c|}{Bias}&\multicolumn{1}{c|}{CI Coverage}\\ \hline
$\Delta(t)$     &0.004            &0.962            & 0.002           &0.954            \\ 
$\Deltastt$     &0.002            &0.971            & 0.013           &0.933            \\ 
$R_{\bS}(t,t_0)$&0.003            &0.955            &-0.048           &0.919            \\ 
\hline
&\multicolumn{4}{c|}{Setting 3}\\ \hline
&\multicolumn{2}{c|}{Plug-in Estimator}&\multicolumn{2}{c|}{TML Estimator}\\ \hline
&\multicolumn{1}{c|}{Bias}&\multicolumn{1}{c|}{CI Coverage}&\multicolumn{1}{c|}{Bias}&\multicolumn{1}{c|}{CI Coverage}\\ \hline
$\Delta(t)$     & 0.003           &0.968            & 0.006           &0.914            \\ 
$\Deltastt$     &-0.007           &0.961            & 0.006           &0.930            \\ 
$R_{\bS}(t,t_0)$& 0.030           &0.947            &-0.010           &0.938            \\ 
  \hline
\end{tabular}
\vspace{3mm}
\end{center}
\end{table}

\section{Example: Diabetes Prevention Program \label{sec:example}}
We illustrate our proposed procedures using data from the Diabetes Prevention Program (DPP), a randomized clinical trial examining metformin, troglitazone, and lifestyle intervention for the prevention of type 2 diabetes in high-risk adults \citep{DPPOS}. At randomization, participants were randomly assigned to one of the treatment groups or placebo. The primary endpoint was time to diabetes as defined by the protocol at the time of the visit: fasting glucose $\geq$ 140 mg/dL (for visits through 6/23/1997, $\geq$ 126 mg/dL for visits on or after 6/24/2007) or 2-h post challenge glucose $\geq$ 200 mg/dL \citep{american1999diabetes}.

\begin{figure}[htbp]
\centering
\includegraphics[scale = 0.7]{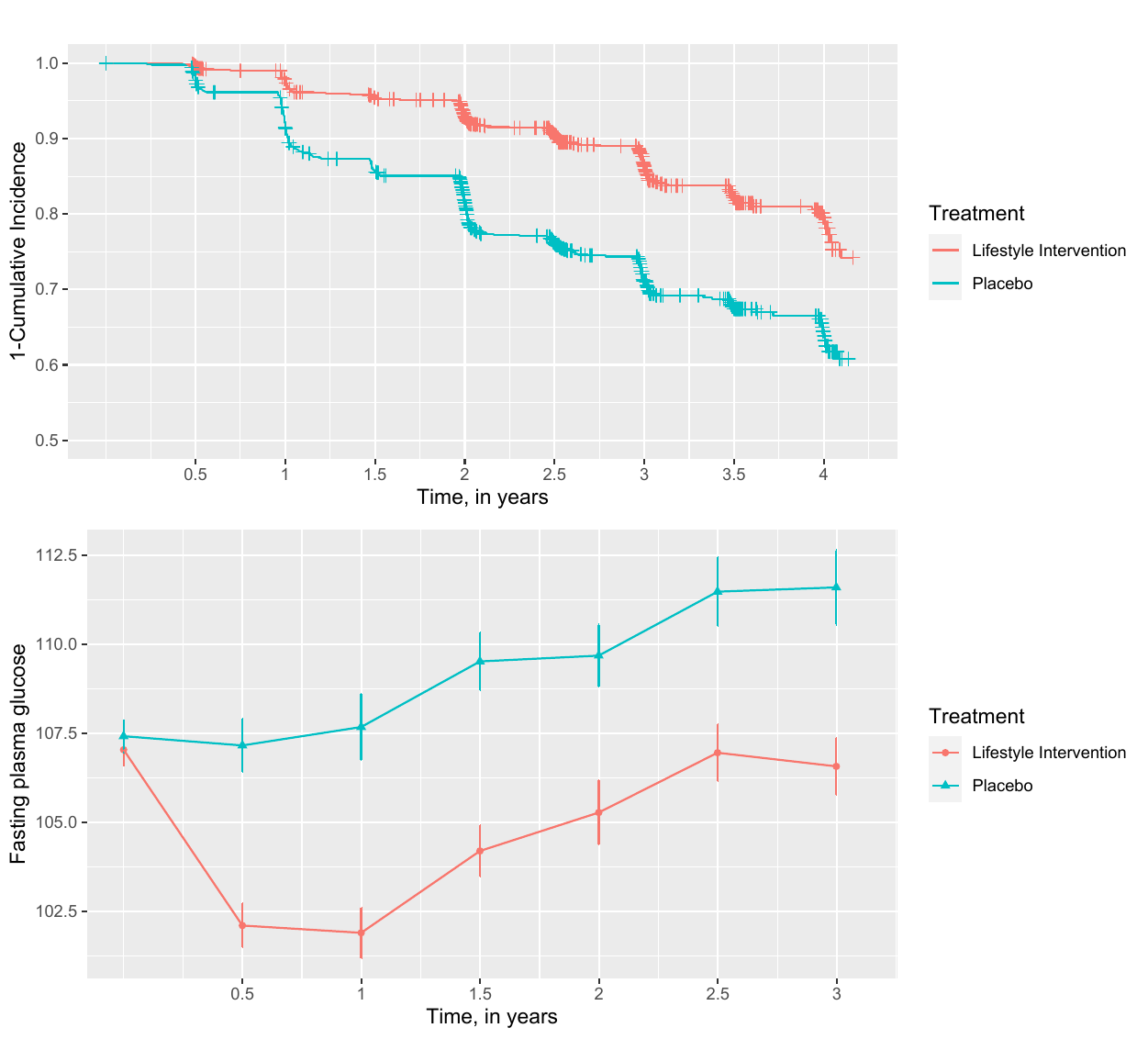}
\caption{Kaplan-Meier estimate of 1 minus the cumulative incidence for diabetes by treatment group (upper plot) and the mean (and 95\% confidence interval) of the surrogate marker, fasting plasma glucose, over time by treatment group (lower plot) \label{dpp_figure}}
\end{figure}

For our analysis, we examined the proportion of treatment effect explained by fasting plasma glucose, measured every 6 months from baseline to 3 years, comparing the lifestyle intervention group (N = 1024) versus placebo (N = 1030). We defined the treatment effect of interest as the difference in (1 minus) the cumulative diabetes incidence at 4 years between the two groups. Figure \ref{dpp_figure} shows the Kaplan-Meier estimate of 1 minus the cumulative incidence for diabetes by treatment group (upper plot) and the mean (and 95\% confidence interval) of the surrogate marker, fasting plasma glucose, over time by treatment group (lower plot).

Using our proposed method, the plug-in estimator for $\Delta(t)$, the difference in (1 minus) the cumulative diabetes incidence at 4 years, was 0.155 (standard error [SE] = 0.038, 95\% confidence interval [CI]: 0.081,0.229), indicating lower cumulative incidence for the lifestyle intervention group. We estimated the residual treatment effect $\Deltastt$ to be much smaller: 0.036 (SE = 0.042, 95\% CI: -0.047, 0.120), which yields an estimate for $R_{\bS}(t,t_0)$ of 0.766 (SE = 0.192, 95\% CI: 0.389, 1.14), suggesting that fasting plasma glucose explains more than 76\% of the treatment effect. Estimation results were somewhat similar using TML: $\Deltaht$ was 0.142 (SE = 0.038, 95\% CI: 0.068, 0.216), $\Deltastt$ was 0.056 (SE = 0.043, 95\% CI: -0.028, 0.140), and $R_{\bS}(t,t_0)$ was 0.605 (SE = 0.178, 95\% CI: 0.257, 0.953). For comparison, though not estimating the same quantity, the estimated proportion of treatment effect explained by the longitudinal surrogate using a joint modeling approach was 0.65 (implemented using the R package \texttt{JMbayes}). Overall, these results indicate that while fasting plasma glucose information up to 3 years captures 60-76\% (depending on the estimator) of the treatment effect on cumulative diabetes incidence at 4 years, this is likely not large enough to be deemed a strong surrogate marker for diabetes, particularly given the low lower bound of the associated confidence intervals.

 \section{Discussion}
We have proposed a robust method for estimating the utility of a surrogate marker in a very general setting with a longitudinal surrogate and a time-to-event outcome. This approach builds on previous work for longitudinal and multivariate surrogate markers \citep{agniel2021evaluation, agniel2020doubly}, and it offers an alternative to parametric methods based on joint modeling or the Cox model. For simplicity of notation, we only included covariates to control for confounding of the average treatment effect on the primary outcome, $\bX$. If additional (longitudinal) covariates are required to justify the independent censoring assumption (Assumption \ref{asmp:indcens2}) or censoring positivity (Assumption \ref{asmp:censpos2}), these may be included at the expense of additional book-keeping throughout.
 
While our approach is quite general, it assumes that the surrogate is collected up until a fixed known time $t_0$. In many cases, the surrogate itself may be expensive to collect, even if it may be less expensive to collect than the gold standard outcome. Even when expense is no issue, collecting the surrogate for a shorter length of time would help speed up future trials. In these cases, selecting the earliest possible $t_0$ could pay dividends in terms of reducing study time and cost. The methods we develop here could be adapted to make such a selection. To this end, let $t_\Lsc < t$ be the latest possible timepoint that would be considered for $t_0$. One approach to consider for selecting $t_0$ is to find the earliest $t^*$ such that the proportion of the treatment effect explained has not declined substantially from $t_\Lsc$. That is, one could select $t_0$ to be the minimum $t^* \in \Omega = \{1,2, ..., t_\Lsc\}$ such that, for a given $\epsilon > 0, R_\bS(t, t^*) > R_\bS(t, t_\Lsc) - \epsilon$. In Appendix \ref{app:test} we sketch out such an approach and use our proposed estimates to describe a procedure to identify the optimal $t_0$ based on the the expected power to detect a treatment effect using the surrogate marker trajectory up to time $t_0$. 

We also note the distinction between our methods here for longitudinal surrogate markers and methods that have been proposed for longitudinal mediation. Certainly, there is a strong connection between methods in these two research areas \citep{vanderweele2013surrogate}. For example, \citet{wang2023targeted} utilize the highly adaptive lasso and projection representations to address longitudinal mediation problems in a causal framework. \citet{lin2017mediation} and \citet{vanderweele2017mediation} propose causal mediation methods based on the g-formula applicable to settings with time-varying exposures and mediators. There are some technical distinctions between our approach here and these alternative approaches: we focus on a binary treatment, while \citet{wang2023targeted} consider more general treatment regimes and target different causal estimands, and \citet{lin2017mediation} and \citet{vanderweele2017mediation} consider different estimands and generally rely on parametric models for estimation. And our approach does not have the difficulties of interpretation or identifiability of mediation estimands in these settings \citep{didelez2019defining}. More importantly, though, the ultimate substantive goal in surrogate marker evaluation is distinct from the goal in mediation. In surrogate marker evaluation, the goal is to \textit{replace} the primary outcome with the surrogate marker in future studies. In general, this is not the goal in mediation. For example, a mediation analysis examining whether income mediates the relationship between race/ethnicity on health outcomes would never imply that income can be used to replace a health outcome in a future study. Mediation focuses solely on the mechanism of the effect and the mediator must be on the causal pathway. A surrogate maker, in some sense, is less restrictive and does not directly make any claim about the mechanism of a treatment effect nor does it necessarily require a surrogate to be on the causal pathway. 

 \section*{Acknowledgement}
 Support for this research was provided by National Institutes of Health grant R01DK118354. The DPP was conducted by the DPP Research Group and supported by the National Institute of Diabetes and Digestive and Kidney Diseases (NIDDK), the General Clinical Research Center Program, the National Institute of Child Health and Human Development (NICHD), the National Institute on Aging (NIA), the Office of Research on Women’s Health, the Office of Research on Minority Health, the Centers for Disease Control and Prevention (CDC), and the American Diabetes Association. The data from the DPP were supplied by the NIDDK Central Repositories. This manuscript was not prepared under the auspices of the DPP and does not represent analyses or conclusions of the DPP Research Group, the NIDDK Central Repositories, or the NIH.

\section*{Supplementary material}
Supplementary material available at Biometrika online includes all references appendices which contain identification and influence function proofs, algorithmic details for the targeted maximum likelihood estimator, further simulation details, and an approach to identify the optimal $t_0$ based on the the expected power to detect a treatment effect using the surrogate marker trajectory up to time $t_0$. All proposed methods are implemented in the R package \texttt{survivalsurrogate} available at \url{https://github.com/denisagniel/survivalsurrogate}.

\bibliographystyle{unsrtnat}
\bibliography{survival-long.bib}

\clearpage
\appendix

\section*{APPENDIX}

\section{Identification and Influence Functions: Proofs for Section \ref{simplified-setting-w-censoring}}\label{app:simplified-setting}

\allowdisplaybreaks
\subsection{Identification of $\Deltat$}\label{app:ident-deltat}
Recall that $\Deltat = \P(T\supA > t) - \P(T\supB > t)$. Due to randomization (assumption \ref{asmp:rand}) and consistency (assumption \ref{asmp:consistency}), $\Deltat = \E(Y_2 | G = 1) - \E(Y_2 | G = 0)$, but $Y_2$ is not observed for all individuals. However,
\begin{align*}
    \E(Y_2 | G = g) &= \E(Y_2 | G = g, Y_1 = 1) \E(Y_1 | G = g).
\end{align*}
Assumptions \ref{asmp:indcens} and \ref{asmp:censpos} ensure that $\E(Y_1 | G = g) = \E(Y_1 | G = g, A_1 = 1)$ and $\E(Y_2 | G = g, Y_1 = 1) = \E\{\E(Y_2 | G = g, A_1 = Y_1 = A_2 = 1, S_1)\},$ yielding 
\begin{align*}
    \E(Y_2 | G = g) &= \E\{\E(Y_2 | G = g, A_1 = Y_1 = A_2 = 1, S_1)\} \E(Y_1 | G = g, A_1 = 1)\\
    &= \mu_{g1}\E\{\mu_{g2}(S_1) | G = g, A_1 = 1\}\\
    &= \int \theta_g(s)dF_{S_1|G = g, A_1 = 1, Y_1 = 1}(s)
\end{align*}
To show the final equality, which gives a weighting-type identification, recall that
\begin{align*}
    \gamma_{g1} &= \frac{f_{GA_1}(g, 1)}{f_{G}(g)}, g = 0, 1\\
    \gamma_{g2}(s) &= \frac{f_{GA_1Y_1S_1A_2}(g, 1, 1, s, 1)}{f_{GA_1Y_1S_1}(g, 1, 1, s)}, g = 0, 1
\end{align*}
so we may write
\begin{align*}
    &\E\left\{\frac{GA_1Y_1A_2Y_2f_{G}(1)f_{GA_1Y_1S_1}(1, 1, 1, S_1)}{f_{GA_1}(1, 1)f_{GA_1Y_1S_1A_2}(1, 1, 1, S_1, 1)} - n/n_0\frac{(1-G)A_1Y_1A_2Y_2f_{G}(0)f_{GA_1Y_1S_1}(0, 1, 1, S_1)}{f_{GA_1}(0, 1)f_{GA_1Y_1S_1A_2}(0, 1, 1, S_1, 1)} \right\}\\
    &=  \int \left\{n/n_1\frac{g a_1y_1a_2y_2 f_{G}(1)f_{GA_1Y_1S_1}(1, 1, 1, s_1)}{f_{GA_1}(1, 1)f_{GA_1Y_1S_1A_2}(1, 1, 1, s_1, 1)} - n/n_0\frac{(1-g)a_1y_1a_2y_2f_{G}(0)f_{GA_1Y_1S_1}(0, 1, 1, s_1)}{f_{GA_1}(0, 1)f_{GA_1Y_1S_1A_2}(0, 1, 1, s_1, 1)} \right\}\times \\
    &\qquad f_{GA_1Y_1S_1A_2Y_2}(g, a_1, y_1, s_1, a_2, y_2)dy_2da_2ds_1dy_1da_1 dg
\end{align*}
Now, the first term is 0 anytime $g = 0, a_1 = 0, y_1 = 0, a_2 = 0$ and $f_G(1) = n_1/n$, so 
\begin{align*}
&\int n/n_1\frac{g a_1y_1a_2y_2 f_{G}(1)f_{GA_1Y_1S_1}(1, 1, 1, s_1)}{f_{GA_1}(1, 1)f_{GA_1Y_1S_1A_2}(1, 1, 1, s_1, 1)}f_{GA_1Y_1S_1A_2Y_2}(g, a_1, y_1, s_1, a_2, y_2)dy_2da_2ds_1dy_1da_1 dg\\
&\qquad = \int\frac{f_{GA_1Y_1S_1}(1, 1, 1, s_1)}{f_{GA_1}(1, 1)}\left\{\int y_2 \frac{f_{GA_1Y_1S_1A_2Y_2}(1, 1, 1, s_1, 1, y_2)}{f_{GA_1Y_1S_1A_2}(1, 1, 1, s_1, 1)}dy_2\right\}ds_1\\
&\qquad= \frac{f_{GA_1Y_1}(1, 1, 1)}{f_{GA_1}(1, 1)}\int\frac{f_{GA_1Y_1S_1}(1, 1, 1, s_1)}{f_{GA_1Y_1}(1, 1, 1)}\mu_{12}(s_1)ds_1\\
&=\mu_{11}\E\{\mu_{12}(S_1) | G = g, A_1 = 1\}
\end{align*}
A similar argument for the second term yields the result. 

\subsection{Identification for $\Deltastt$}\label{app:ident-deltastt}
Following a similar argument as above, due to randomization (assumption \ref{asmp:rand}), positivity (assumption \ref{asmp:pos}), and consistency (assumption \ref{asmp:consistency}), $\P(T\supg > t | S_1\supg = s, T\supg > 1) = \E(Y_2 | S_1 = s, G = g, Y_1 = 1)$ and $\P(T\supg > 1) = \E(Y_1 | G = g)$. Assumptions \ref{asmp:indcens} and \ref{asmp:censpos} ensure that $\E(Y_1 | G = g) = \E(Y_1 | G = g, A_1 = 1)$,  $\E(Y_2 | S_1 = s, G = g, Y_1 = 1) = \E(Y_2 | G = g, A_1 = Y_1 = A_2 = 1, S_1),$ and $F_{S_1 | T > 1}(s) = F_{S_1|A_1=1,Y_1 = 1}(s)$, which yields the result:
\begin{align*}
    \Deltastt &= \int \left\{\P(T\supA > t | S_1\supA = s, T\supA > 1)\P(T\supA > 1) - \right.\\
    &\qquad\qquad \left.\P(T\supB > t | S_1\supB = s, T\supB > 1)\P(T\supB > 1)\right\}dF_{S_1 | T > 1}(s)\\
    &= \int \left\{\E(Y_2 | S_1 = s, G = 1, A_1 = Y_1 = A_2 = 1)\E(Y_1 | G = 1, A_1 = 1) - \right.\\
    &\qquad\qquad \left.\E(Y_2 | S_1 = s, G = 0, A_1 = Y_1 = A_2 = 1)\E(Y_1 | G = 0, A_1 = 1)\right\}dF_{S_1 | A_1 = 1, Y_1 =1}(s)\\
    &= \int \left\{\theta_1(s) - \theta_0(s)\right\}dF_{S_1 | A_1 = 1, Y_1 =1}(s).
\end{align*}

To show the final equality, recall that
\begin{align*}
    \gamma_{g1} &= \frac{f_{GA_1}(g, 1)}{f_{G}(g)}, g = 0, 1\\
    \gamma_{g2}(s) &= \frac{f_{GA_1Y_1S_1A_2}(g, 1, 1, s, 1)}{f_{GA_1Y_1S_1}(g, 1, 1, s)}, g = 0, 1\\
    \pi_1(s) &= \frac{f_{GA_1Y_1S_1}(1, 1, 1, s)}{f_{A_1Y_1S_1}(1, 1, s)}\\
    \pi_1^* &= \frac{f_{GA_1Y_1}(1, 1, 1)}{f_{A_1Y_1}(1, 1)},
\end{align*}
so we may write
\begin{align*}
& \E\left\{n/n_1\frac{GA_1Y_1A_2Y_2f_{G}(1)}{f_{GA_1}(1, 1)f_{GA_1Y_1S_1A_2}(1, 1, 1, s, 1)} \times \frac{f_{A_1Y_1S_1}(1, 1, s)f_{GA_1Y_1}(1, 1, 1)}{f_{A_1Y_1}(1, 1)} -\right.\\ &\left.\qquad n/n_0\frac{(1-G)A_1Y_1A_2Y_2f_{G}(0)}{f_{GA_1}(0, 1)f_{GA_1Y_1S_1A_2}(0, 1, 1, s, 1)} \times \frac{f_{A_1Y_1S_1}(1, 1, s)f_{GA_1Y_1}(0, 1, 1)}{f_{A_1Y_1}(1, 1)} \right\}\\
    &= \int \left\{n/n_1\frac{ga_1y_1a_2y_2f_{G}(1)}{f_{GA_1}(1, 1)f_{GA_1Y_1S_1A_2}(1, 1, 1, s, 1)} \times \frac{f_{A_1Y_1S_1}(1, 1, s)f_{GA_1Y_1}(1, 1, 1)}{f_{A_1Y_1}(1, 1)} -\right.\\ &\left.\qquad n/n_0\frac{(1-g)a_1y_1a_2y_2f_{G}(0)}{f_{GA_1}(0, 1)f_{GA_1Y_1S_1A_2}(0, 1, 1, s, 1)} \times \frac{f_{A_1Y_1S_1}(1, 1, s)f_{GA_1Y_1}(0, 1, 1)}{f_{A_1Y_1}(1, 1)} \right\}\times\\
    &\qquad f_{GA_1Y_1S_1A_2Y_2}(g, a_1, y_1, s, a_2, y_2)dgda_1dy_1dsda_2dy_2\\
    &= \int \left\{\frac{f_{GA_1Y_1}(1, 1, 1)}{f_{GA_1}(1, 1)} \times \frac{f_{GA_1Y_1S_1A_2Y_2}(1, 1, 1, s, 1, 1)}{f_{GA_1Y_1S_1A_2}(1, 1, 1, s, 1)} -\right. \\ 
    &\qquad \left.\frac{f_{GA_1Y_1}(0, 1, 1)}{f_{GA_1}(0, 1)} \times \frac{f_{GA_1Y_1S_1A_2Y_2}(0, 1, 1, s, 1, 1)}{f_{GA_1Y_1S_1A_2}(0, 1, 1, s, 1)}\right\} \frac{f_{A_1Y_1S_1}(1, 1, s)}{f_{A_1Y_1}(1, 1)}ds\\
    &= \int \left\{\theta_1(s) - \theta_0(s)\right\}dF_{S_1 | A_1 = 1, Y_1 =1}(s).
\end{align*}

\subsection{Influence function for $\Deltat$ \label{if_delta_section}}
Note that 
\begin{align*}
    \Deltat = \E\left\{n/n_1\frac{GA_1Y_1A_2Y_2}{\gamma_{11}\gamma_{12}(S_1)} - n/n_0\frac{(1-G)A_1Y_1A_2Y_2}{\gamma_{01}\gamma_{02}(S_1)} \right\}.
\end{align*}
Recall that
\begin{align*}
    \gamma_{g1} &= \frac{f_{GA_1}(g, 1)}{f_{G}(g)}, g = 0, 1\\
    \gamma_{g2}(s) &= \frac{f_{GA_1Y_1S_1A_2}(g, 1, 1, s, 1)}{f_{GA_1Y_1S_1}(g, 1, 1, s)}, g = 0, 1
\end{align*}
so we may rewrite $\Deltat$ as
\begin{align*}
    \Deltat &= \E\left\{n/n_1\frac{GA_1Y_1A_2Y_2f_{G}(1)f_{GA_1Y_1S_1}(1, 1, 1, s)}{f_{GA_1}(1, 1)f_{GA_1Y_1S_1A_2}(1, 1, 1, s, 1)} - n/n_0\frac{(1-G)A_1Y_1A_2Y_2f_{G}(0)f_{GA_1Y_1S_1}(0, 1, 1, s)}{f_{GA_1}(0, 1)f_{GA_1Y_1S_1A_2}(0, 1, 1, s, 1)} \right\}
\end{align*}
Letting the observed data distribution be $\Psc$, we can express $\Deltat$ as 
\begin{align*}
    \Deltat &= n/n_1\Psi_1(\Psc) - n/n_0\Psi_0(\Psc)\\
    \Psi_1(\Psc) &= \E_\Psc\left\{\frac{GA_1Y_1A_2Y_2f_{G}(1)f_{GA_1Y_1S_1}(1, 1, 1, s)}{f_{GA_1}(1, 1)f_{GA_1Y_1S_1A_2}(1, 1, 1, s, 1)}\right\}\\
    \Psi_0(\Psc) &= \E_\Psc\left\{\frac{(1-G)A_1Y_1A_2Y_2f_{G}(0)f_{GA_1Y_1S_1}(0, 1, 1, s)}{f_{GA_1}(0, 1)f_{GA_1Y_1S_1A_2}(0, 1, 1, s, 1)}\right\}.
\end{align*}
where we make the dependence of the expectation on $\Psc$ explicit. To find the influence function for $\Psi_g(\Psc)$, we compute $d/d\epsilon \Psi_g(\Psc_\epsilon)$ for the mixture model $\Psc_\epsilon = \epsilon\Psctilde + (1-\epsilon)\Psc$, where $\Psctilde$ is a point mass at $\otilde = (\gtilde, \atilde_1, \ytilde_1, \stilde, \atilde_2, \ytilde_2)$.
\begin{align*}
    \frac{d}{d\epsilon}\Psi_1(\Psc_\epsilon) &= \int \frac{ga_1y_1a_2y_2f_{G}(1)f_{GA_1Y_1S_1}(1,1, 1, s)}{f_{GA_1}(1, 1)f_{GA_1Y_1S_1A_2}(1, 1, 1, s, 1)} \times\\ 
    &\qquad \left\{I_{GA_1Y_1S_1A_2Y_2}(\gtilde, \atilde_1, \ytilde_1, \stilde, \atilde_2, \ytilde_2) - f_{GA_1Y_1S_1A_2Y_2}(g, a_1, y_1, s, a_2, y_2)\right\}dgda_1dy_1dsda_2dy_2 - \\
    &\qquad \int \frac{ga_1y_1a_2y_2f_{G}(1)f_{GA_1Y_1S_1}(1, 1, 1, s)}{f_{GA_1}(1, 1)f_{GA_1Y_1S_1A_2}(1, 1, 1, s, 1)^2}f_{GA_1Y_1S_1A_2Y_2}(g, a_1, y_1, s, a_2, y_2) \times\\ 
    &\qquad \left\{I_{GA_1Y_1S_1A_2}(\gtilde, \atilde_1, \ytilde_1, \stilde, \atilde_2) - f_{GA_1Y_1S_1A_2}(g, a_1, y_1, s, a_2)\right\}dgda_1dy_1dsda_2dy_2 + \\
    &\qquad \int \frac{ga_1y_1a_2y_2f_{G}(1)}{f_{GA_1}(1, 1)f_{GA_1Y_1S_1A_2}(1, 1, 1, s, 1)}f_{GA_1Y_1S_1A_2Y_2}(g, a_1, y_1, s, a_2, y_2) \times\\ 
    &\qquad \left\{I_{GA_1Y_1S_1}(\gtilde, \atilde_1, \ytilde_1, \stilde) - f_{GA_1Y_1S_1}(g, a_1, y_1, s)\right\}dgda_1dy_1dsda_2dy_2 +\\
    &\qquad \int \frac{ga_1y_1a_2y_2f_{G}(1)f_{GA_1Y_1S_1}(1,1, 1, s)}{f_{GA_1}(1, 1)^2f_{GA_1Y_1S_1A_2}(1, 1, 1, s, 1)}f_{GA_1Y_1S_1A_2Y_2}(g, a_1, y_1, s, a_2, y_2) \times\\ 
    &\qquad \left\{I_{GA_1}(\gtilde, \atilde_1) - f_{GA_1}(g, a_1)\right\}dgda_1dy_1dsda_2dy_2+\\
    &\qquad \int \frac{ga_1y_1a_2y_2f_{GA_1Y_1S_1}(1,1, 1, s)}{f_{GA_1}(1, 1)f_{GA_1Y_1S_1A_2}(1, 1, 1, s, 1)}f_{GA_1Y_1S_1A_2Y_2}(g, a_1, y_1, s, a_2, y_2) \times\\ 
    &\qquad \left\{I_{G}(\gtilde) - f_{G}(g)\right\}dgda_1dy_1dsda_2dy_2\\
    &= I + II + III + IV + V
\end{align*}
Taking the five terms in turn, the first term yields a centered version of the typical IPW estimator:
\begin{align*}
    I = \frac{\gtilde \atilde_1 \ytilde_1 \atilde_2 \ytilde_2}{\gamma_{11}\gamma_{12}(\stilde)} - \Psi_1(\Psc).
\end{align*}
The second term yields a similarly weighted version of the mean function
\begin{align*}
    II = -\frac{\gtilde \atilde_1 \ytilde_1 \atilde_2\mu_1(1, \stilde)}{\gamma_{11}\gamma_{12}(\stilde)} + \Psi_1(\Psc).
\end{align*}

\begin{align*}
    III &= \gtilde\atilde_1 \ytilde_1\int \frac{a_2y_2f_G(1)}{f_{GA_1}(1, 1)f_{GA_1Y_1S_1A_2}(1, 1, 1, \stilde, 1)}f_{GA_1Y_1S_1A_2Y_2}(\gtilde, \atilde_1, \ytilde_1, \stilde_1, a_2, y_2) da_2dy_2 - \Psi_1(\Psc)\\
    &= \frac{\gtilde\atilde_1\ytilde_1}{\gamma_{11}}\mu_1(1, \stilde) - \Psi_1(\Psc).
\end{align*}
\begin{align*}
    IV = -\frac{\gtilde \atilde_1}{\gamma_{11}}\E\{\mu_{12}(S_1) | G = 1, A_1 = 1, Y_1 = 1\}\E(Y_1 | G = 1, A_1 = 1) + \Psi_1(\Psc)
\end{align*}

\begin{align*}
    V = \gtilde\E\{\mu_{12}(S_1) | G = 1, A_1 = 1, Y_1 = 1\}\E(Y_1 | G = 1, A_1 = 1) - \Psi_1(\Psc)
\end{align*}
And again the result in the main text by following this same process for $\Psi_0$.

\subsection{Influence function for $\Deltastt$ \label{if_deltas_section}}
As above, we may rewrite $\Deltastt$ as
\begin{align*}
    \Deltastt &=  \int \left\{n/n_1\frac{ga_1y_1a_2y_2f_{G}(1)}{f_{GA_1}(1, 1)f_{GA_1Y_1S_1A_2}(1, 1, 1, s, 1)} \times \frac{f_{A_1Y_1S_1}(1, 1, s)f_{GA_1Y_1}(1, 1, 1)}{f_{A_1Y_1}(1, 1)} -\right.\\ &\left.\qquad n/n_0\frac{(1-g)a_1y_1a_2y_2f_{G}(0)}{f_{GA_1}(0, 1)f_{GA_1Y_1S_1A_2}(0, 1, 1, s, 1)} \times \frac{f_{A_1Y_1S_1}(1, 1, s)f_{GA_1Y_1}(0, 1, 1)}{f_{A_1Y_1}(1, 1)} \right\}\times\\
    &\qquad f_{GA_1Y_1S_1A_2Y_2}(g, a_1, y_1, s, a_2, y_2)dgda_1dy_1dsda_2dy_2
\end{align*}
Letting the observed data distribution be $\Psc$, we can express $\Deltastt$ as 
\begin{align*}
    \Deltastt &= n/n_1\Psi_1(\Psc) - n/n_0\Psi_0(\Psc)\\
    \Psi_1(\Psc) &= \E_\Psc\left\{\frac{GA_1Y_1A_2Y_2f_{G}(1)}{f_{GA_1}(1, 1)f_{GA_1Y_1S_1A_2}(1, 1, 1, s, 1)} \times \frac{f_{A_1Y_1S_1}(1, 1, s)f_{GA_1Y_1}(1, 1, 1)}{f_{A_1Y_1}(1, 1)}\right\}\\
    \Psi_0(\Psc) &= \E_\Psc\left\{\frac{(1-G)A_1Y_1A_2Y_2f_{G}(0)}{f_{GA_1}(0, 1)f_{GA_1Y_1S_1A_2}(0, 1, 1, s, 1)} \times \frac{f_{A_1Y_1S_1}(1, 1, s)f_{GA_1Y_1}(0, 1, 1)}{f_{A_1Y_1}(1, 1)}\right\}.
\end{align*}
where we make the dependence of the expectation on $\Psc$ explicit. To find the influence function for $\Psi_g(\Psc)$, we compute $d/d\epsilon \Psi_g(\Psc_\epsilon)$ for the mixture model $\Psc_\epsilon = \epsilon\Psctilde + (1-\epsilon)\Psc$, where $\Psctilde$ is a point mass at $\otilde = (\gtilde, \atilde_1, \ytilde_1, \stilde, \atilde_2, \ytilde_2)$.
\begin{align*}
    \frac{d}{d\epsilon}\Psi_1(\Psc_\epsilon) &= \int \frac{ga_1y_1a_2y_2f_{G}(1)f_{A_1Y_1S_1}(1, 1, s)f_{GA_1Y_1}(1, 1, 1)}{f_{GA_1}(1, 1)f_{GA_1Y_1S_1A_2}(1, 1, 1, s, 1)f_{A_1Y_1}(1, 1)} \times\\ &\qquad \left\{I_{GA_1Y_1S_1A_2Y_2}(\gtilde, \atilde_1, \ytilde_1, \stilde, \atilde_2, \ytilde_2) - f_{GA_1Y_1S_1A_2Y_2}(g, a_1, y_1, s, a_2, y_2)\right\}dgda_1dy_1dsda_2dy_2 - \\
    &\qquad \int \frac{ga_1y_1a_2y_2f_{G}(1)f_{A_1Y_1S_1}(1, 1, s)f_{GA_1Y_1}(1, 1, 1)}{f_{GA_1}(1, 1)f_{GA_1Y_1S_1A_2}(1, 1, 1, s, 1)^2f_{A_1Y_1}(1, 1)}f_{GA_1Y_1S_1A_2Y_2}(g, a_1, y_1, s, a_2, y_2) \times\\ &\qquad \left\{I_{GA_1Y_1S_1A_2}(\gtilde, \atilde_1, \ytilde_1, \stilde, \atilde_2) - f_{GA_1Y_1S_1A_2}(g, a_1, y_1, s, a_2)\right\}dgda_1dy_1dsda_2dy_2 + \\
    &\qquad \int \frac{ga_1y_1a_2y_2f_{G}(1)f_{GA_1Y_1}(1, 1, 1)}{f_{GA_1}(1, 1)f_{GA_1Y_1S_1A_2}(1, 1, 1, s, 1)f_{A_1Y_1}(1, 1)}f_{GA_1Y_1S_1A_2Y_2}(g, a_1, y_1, s, a_2, y_2) \times\\ 
    &\qquad \left\{I_{A_1Y_1S_1}( \atilde_1, \ytilde_1, \stilde) - f_{A_1Y_1S_1}(a_1, y_1, s)\right\}dgda_1dy_1dsda_2dy_2 +\\
    &\qquad \int \frac{ga_1y_1a_2y_2f_{G}(1)f_{A_1Y_1S_1}(a_1, y_1, s)}{f_{GA_1}(1, 1)f_{GA_1Y_1S_1A_2}(1, 1, 1, s, 1)f_{A_1Y_1}(1, 1)}f_{GA_1Y_1S_1A_2Y_2}(g, a_1, y_1, s, a_2, y_2) \times\\ 
    &\qquad \left\{I_{GA_1Y_1}(\gtilde, \atilde_1, \ytilde_1) - f_{GA_1Y_1}(1, 1, 1)\right\}dgda_1dy_1dsda_2dy_2 -\\
    &\qquad \int \frac{ga_1y_1a_2y_2f_{G}(1)f_{A_1Y_1S_1}(1, 1, s)f_{GA_1Y_1}(1, 1, 1)}{f_{GA_1}(1, 1)f_{GA_1Y_1S_1A_2}(1, 1, 1, s, 1)f_{A_1Y_1}(1, 1)^2}f_{GA_1Y_1S_1A_2Y_2}(g, a_1, y_1, s, a_2, y_2) \times\\ 
    &\qquad \left\{I_{A_1Y_1}(\atilde_1, \ytilde_1) - f_{A_1Y_1}(a_1, y_1)\right\}dgda_1dy_1dsda_2dy_2 -\\
    &\qquad \int \frac{ga_1y_1a_2y_2f_{G}(1)f_{A_1Y_1S_1}(1, 1, s)f_{GA_1Y_1}(1, 1, 1)}{f_{GA_1}(1, 1)^2f_{GA_1Y_1S_1A_2}(1, 1, 1, s, 1)f_{A_1Y_1}(1, 1)}f_{GA_1Y_1S_1A_2Y_2}(g, a_1, y_1, s, a_2, y_2) \times\\ 
    &\qquad \left\{I_{GA_1}(\gtilde, \atilde_1) - f_{GA_1}(g, a_1)\right\}dgda_1dy_1dsda_2dy_2+\\
    &\qquad \int \frac{ga_1y_1a_2y_2f_{A_1Y_1S_1}(1, 1, s)f_{GA_1Y_1}(1, 1, 1)}{f_{GA_1}(1, 1)f_{GA_1Y_1S_1A_2}(1, 1, 1, s, 1)f_{A_1Y_1}(1, 1)}f_{GA_1Y_1S_1A_2Y_2}(g, a_1, y_1, s, a_2, y_2) \times\\ 
    &\qquad \left\{I_{G}(\gtilde) - f_{G}(g)\right\}dgda_1dy_1dsda_2dy_2\\
    &= I + II + III + IV + V + VI + VII
\end{align*}
Taking the terms in turn, the first term yields a centered version of the typical IPW estimator:
\begin{align*}
    I = \frac{\gtilde \atilde_1 \ytilde_1 \atilde_2 \ytilde_2f_G(1) f_{A_1Y_1S_1}(1, 1, \stilde)f_{GA_1Y_1}(1, 1, 1)}{f_{GA_1}(1, 1)f_{GA_1Y_1S_1A_2}(1, 1, 1, \stilde, 1)f_{A_1Y_1}(1, 1)} - \Psi_1(\Psc) = \frac{\gtilde \atilde_1 \ytilde_1 \atilde_2 \ytilde_2}{\gamma_{11}\gamma_{12}(\stilde)} \times \frac{\pi^*_1}{\pi_1(\stilde)} - \Psi_1(\Psc).
\end{align*}
The second term yields a similarly weighted version of the mean function
\begin{align*}
    II &= -\int \frac{\gtilde \atilde_1 \ytilde_1 \atilde_2 y_2f_G(1)f_{A_1Y_1S_1}(1, 1, \stilde)f_{GA_1Y_1}(1, 1, 1)}{f_{GA_1}(1, 1)f_{GA_1Y_1S_1A_2}(1, 1, 1, \stilde, 1)^2f_{A_1Y_1}(1, 1)}f_{GA_1Y_1S_1A_2Y_2}(\gtilde, \atilde_1, \ytilde_1, \stilde, \atilde_2, y_2)dy_2 + \Psi_1(\Psc)\\
    &= -\frac{\gtilde \atilde_1 \ytilde_1 \atilde_2}{\gamma_{11}\gamma_{12}(\stilde)} \times \frac{\pi^*_1}{\pi_1(\stilde)} \times \frac{f_{GA_1Y_1S_1A_2}(\gtilde, \atilde_1, \ytilde_1, \stilde, \atilde_2)}{f_{GA_1Y_1S_1A_2}(1, 1, 1, \stilde, 1)} \times \E(Y_2 | G = \gtilde, A_1 = \atilde_1, Y_1 = \ytilde_1, S_1 = \stilde, A_2 = \atilde_2) + \Psi_1(\Psc).\\
    &= -\frac{\gtilde \atilde_1 \ytilde_1 \atilde_2\mu_1(1, \stilde)}{\gamma_{11}\gamma_{12}(\stilde)} \times \frac{\pi^*_1}{\pi_1(\stilde)} + \Psi_1(\Psc).
\end{align*}

\begin{align*}
    III &= \atilde_1 \ytilde_1\int \frac{ga_2y_2f_G(1)f_{GA_1Y_1}(1, 1, 1)}{f_{GA_1}(1, 1)f_{GA_1Y_1S_1A_2}(1, 1, 1, \stilde, 1)f_{A_1Y_1}(1, 1)}f_{GA_1Y_1S_1A_2Y_2}(g, \atilde_1, \ytilde_1, \stilde_1, a_2, y_2) dgda_2dy_2 - \Psi_1(\Psc)\\
    &= \frac{\atilde_1\ytilde_1\pi_1^*}{\gamma_{11}}\mu_1(1, \stilde) - \Psi_1(\Psc).
\end{align*}

\begin{align*}
    IV &= \gtilde \atilde_1 \ytilde_1\int \frac{a_2y_2f_G(1)f_{A_1Y_1S_1}(1, 1, s)}{f_{GA_1}(1, 1)f_{GA_1Y_1S_1A_2}(1, 1, 1, s, 1)f_{A_1Y_1}(1, 1)}f_{GA_1Y_1S_1A_2Y_2}(\gtilde, \atilde_1, \ytilde_1, s, a_2, y_2)dsda_2dy_2 - \Psi_1(\Psc)\\
    &= \frac{\gtilde\atilde_1\ytilde_1}{\gamma_{11}}\E\{\mu_1(1, S) | A_1 = 1, Y_1 = 1\}  - \Psi_1(\Psc)
\end{align*}

\begin{align*}
    V &= -\int \frac{g \atilde_1 \ytilde_1 a_2y_2f_{G}(1)f_{A_1Y_1S_1}(1, 1, s)f_{GA_1Y_1}(1, 1, 1)}{f_{GA_1}(1, 1)f_{GA_1Y_1S_1A_2}(1, 1, 1, s, 1)f_{A_1Y_1}(1, 1)^2}f_{GA_1Y_1S_1A_2Y_2}(g, \atilde_1, \ytilde_1, s, a_2, y_2)dgdsda_2dy_2 + \Psi_1(\Psc)\\
    &= -\frac{\atilde_1\ytilde_1\pi_1^*}{\gamma_{11}}\E\{\mu_1(1, S) | A_1 = 1, Y_1 = 1\}  + \Psi_1(\Psc)
\end{align*}

\begin{align*}
    VI &=  -\int \frac{\gtilde \atilde_1 y_1a_2y_2f_{G}(1)f_{A_1Y_1S_1}(1, 1, s)f_{GA_1Y_1}(1, 1, 1)}{f_{GA_1}(1, 1)^2f_{GA_1Y_1S_1A_2}(1, 1, 1, s, 1)f_{A_1Y_1}(1, 1)}f_{GA_1Y_1S_1A_2Y_2}(\gtilde, \atilde_1, y_1, s, a_2, y_2)dy_1dsda_2dy_2 + \Psi_1(\Psc)\\
    &= -\frac{\gtilde \atilde_1}{\gamma_{11}}\E\{\mu_1(1, S) | A_1 = 1, Y_1 = 1\}\E(Y_1 | G = 1, A_1 = 1) + \Psi_1(\Psc)
\end{align*}

\begin{align*}
    VII &= \int \frac{\gtilde a_1y_1a_2y_2f_{A_1Y_1S_1}(1, 1, s)f_{GA_1Y_1}(1, 1, 1)}{f_{GA_1}(1, 1)f_{GA_1Y_1S_1A_2}(1, 1, 1, s, 1)f_{A_1Y_1}(1, 1)}f_{GA_1Y_1S_1A_2Y_2}(\gtilde, a_1, y_1, s, a_2, y_2)da_1dy_1dsda_2dy_2 - \Psi_1(\Psc)\\
    &= \gtilde\E\{\mu_1(1, S) | A_1 = 1, Y_1 = 1\}\E(Y_1 | G = 1, A_1 = 1) - \Psi_1(\Psc)
\end{align*}
The form of the influence function in the main text follows by repeating this process for $\Psi_0$.

\section{Identification and Influence Functions: Proofs for Section \ref{sec:main-setting}}\label{app:main-setting}

\subsection{Identification for $\Deltat$}\label{app:main-setting-id-deltat}

Recall that $\Deltat = \P(T\supA > t) - \P(T\supB > t)$. we can write each of these terms as
\begin{align*}
    \P(T\supg > t) = \E(Y_t\supg) = \E_\bX\left\{\E(Y_t\supg | \bX, G = g, A_1 = 1)\right\} = \E_\bX\left\{\E(Y_t | \bX, G = g, A_1 = 1)\right\}
\end{align*}
because of assumptions \ref{asmp:consistency}, \ref{asmp:ign}, and \ref{asmp:indcens2}. We may further write
\begin{align}
    \E_\bX\left\{\E(Y_t | \bX, G = g, A_1 = 1)\right\} &= \E_\bX\left\{\E(Y_t | \bX, G = g, A_1 = Y_1 = 1)\E(Y_1 | \bX, G = g, A_1 = 1)\right\}\\
    &= \E_\bX\left[\E_{S_1}\left\{\E(Y_t | \bX, G = g, A_1 = Y_1 = 1, S_1)\right\}\mu_{g1}(\bX)\right]\label{deltat-gcomp}
\end{align}
Because of assumption \ref{asmp:indcens2}, it follows that for $k = 2, ..., t_0$
\begin{align*}
    \E(Y_t | \bX, G = g, \bAbar_{k-1} = \bYbar_{k-1} = 1, \bSbar_{k-1}) &= \E(Y_t | \bX, G = g, \bAbar_k = \bYbar_{k-1} = 1, \bSbar_{k-1})\\
    &= \E(Y_t | \bX, G = g, \bAbar_k = \bYbar_k = 1, \bSbar_{k-1})\mu_{gk}(\bX, \bSbar_{k-1})\\
    &= \E_{S_k}\left\{\E(Y_t | \bX, G = g, \bAbar_k = \bYbar_k = 1, \bSbar_k)\right\}\mu_{gk}(\bX, \bSbar_{k-1}).
\end{align*}
For $k = t_0+1, ..., t - 1$, we have
\begin{align*}
    \E(Y_t | \bX, G = g, \bAbar_{k-1} = \bYbar_{k-1} = 1, \bSbar_{t_0}) &= \E(Y_t | \bX, G = g, \bAbar_k = \bYbar_{k-1} = 1, \bSbar_{t_0})\\
    &= \E(Y_t | \bX, G = g, \bAbar_k = \bYbar_k = 1, \bSbar_{t_0})\mu_{gk}(\bX, \bSbar_{t_0}).
\end{align*}
Plugging these values into \eqref{deltat-gcomp} yields the first identification.

To see the second identification, we can write the second IPW-type identification as 
\begin{align*}
    \Deltat &= \E\left\{ 
                    \frac{G}{e(\bX)} \prod_{k=1}^t \frac{A_{ik}Y_{ik}}{\gamma_{1k}(\bX, \bSbar_{k-1})} - \frac{1-G}{1-e(\bX)} \prod_{k=1}^t \frac{A_{ik}Y_{ik}}{\gamma_{0k}(\bX, \bSbar_{k-1})}
                  \right\}
                  &= \Psi_1 - \Psi_0.
\end{align*}
Looking at the first term,
\begin{align*}
    \Psi_1 &= \int\left\{ 
                    \frac{g}{e(\bx)} \prod_{k=1}^t \frac{a_ky_k}{\gamma_{1k}(\bx, \bsbar_{k-1})}
                  \right\}f_{Y_tA_t\bWbar_{t-1}}(y_t, a_t, \bwbar_{t-1})dy_tda_td\bwbar_{t-1}\\
                  &= \int\left\{ 
                    \frac{gf_\bX(\bx)}{f_{\bX G}(\bx, 1)} \prod_{k=1}^t \frac{a_ky_kf_{\bX G\bAbar_{k-1}\bYbar_{k-1}\bSbar_{k-1}}(\bx, 1, \bone_{k-1}, \bone_{k-1}, \bsbar_{k-1}) }{f_{\bX G\bAbar_k\bYbar_{k-1}\bSbar_{k-1}}(\bx, 1, \bone_{k}, \bone_{k-1} \bsbar_{k-1})}
                  \right\}\times\\
                  &\qquad\qquad f_{\bX G\bAbar_t\bYbar_t\bSbar_{t-1}}(\bx, g, \babar_t, \bybar_t, \bsbar_{t-1})d\bybar_td\babar_td\bsbar_{t-1}dgd\bx\\
                  &= \int\left\{ f_\bX(\bx)
                    \frac{g}{f_{\bX G}(\bx, 1)} \prod_{k=1}^t \frac{a_ky_kf_{\bX G\bAbar_{k-1}\bYbar_{k-1}\bSbar_{k-1}}(\bx, 1, \bone_{k-1}, \bone_{k-1}, \bsbar_{k-1}) }{f_{\bX G\bAbar_k\bYbar_{k-1}\bSbar_{k-1}}(\bx, 1, \bone_{k}, \bone_{k-1} \bsbar_{k-1})}
                  \right\}\times\\
                  &\qquad\qquad f_{\bX G\bAbar_t\bYbar_t\bSbar_{t-1}}(\bx, g, \babar_t, \bybar_t, \bsbar_{t-1})d\bybar_td\babar_td\bsbar_{t-1}dgd\bx\\
                  &= \int f_\bX(\bx) g \times a_1y_1 \frac{f_{\bX G A_1 Y_1 S_1}(\bx, 1, 1, 1, s_1)}{f_{\bX GA_1}(\bx, 1, 1)} \times a_2y_2 \frac{f_{\bX G \bAbar_2 \bYbar_2 \bSbar_2}(\bx, 1, \bone_2, \bone_2, \bsbar_2)}{f_{\bX G\bAbar_2Y_1S_1}(\bx, 1, \bone_2, 1, s_1)} \times ... \times\\
                  &\qquad\qquad  a_ty_t \frac{f_{\bX G \bAbar_t \bYbar_t \bSbar_{t-1}}(\bx, 1, \bone_t, \bone_t, \bsbar_{t-1})}{f_{\bX G\bAbar_tY_{t-1}S_{t-1}}(\bx, 1, \bone_t, \bone_{t-1}, \bsbar_{t-1})}d\bybar_td\babar_td\bsbar_{t-1}dgd\bx\\
                  &= \Qsc_{10}.
\end{align*}
The result in the main text follows from repeating this process for $\Psi_0$.

\subsection{Identification for $\Deltastt$}\label{app:main-setting-id-deltastt}
The identification of $\Deltastt$ follows the same lines as the identification of $\Deltat$. Write  \begin{align*}
    \Deltastt &= \int \left\{\prod_{k=1}^t\P(T\supA > k | T\supA > k - 1, \bSbar\supA_{k-1} = \bsbar_{k-1}) - \right.\\
    &\qquad\qquad \left.\prod_{k=1}^t\P(T\supB > t | T\supB > k - 1, \bSbar\supB_{k-1} = \bsbar_{k-1})\right\}d\left\{\prod_{k=1}^{t_0} F_{S_k | T > k}(s_k)\right\}\\
    &= \int \left\{\prod_{k=1}^t\P(Y_k | \bX = \bx, G = 1, \bYbar_{k-1} = 1, \bSbar_{k-1} = \bsbar_{k-1}) - \right.\\
    &\qquad\qquad \left.\prod_{k=1}^t\P(Y_k | \bX = \bx, G = 0,  \bYbar_{k - 1}=1, \bSbar_{k-1} = \bsbar_{k-1})\right\}d\left\{\prod_{k=1}^{t_0} F_{S_k | T > k}(s_k)\right\}dF_\bX(\bx)\\
\end{align*}
because of assumptions \ref{asmp:consistency} and \ref{asmp:ign}. Further, because of assumption \ref{asmp:indcens2}, we have 
\begin{align*}
     \Deltastt &= \int \left\{\prod_{k=1}^t\P(Y_k | \bX = \bx, G = 1, \bYbar_{k-1} = \bAbar_k = 1, \bSbar_{k-1} = \bsbar_{k-1}) - \right.\\
    &\qquad\qquad \left.\P(Y_k | \bX = \bx, G = 0,  \bYbar_{k - 1} = \bAbar_k=1, \bSbar_{k-1} = \bsbar_{k-1})\right\}d\left\{\prod_{k=1}^{t_0} F_{S_k | T > k}(s_k)\right\}dF_\bX(\bx)\\
    &= \int \left\{\prod_{k=1}^t\mu_{1k}(\bX, \bSbar_{k-1}) - \prod_{k=1}^t\mu_{0k}(\bX, \bSbar_{k-1})\right\}d\left\{\prod_{k=1}^{t_0} F_{S_k | T > k}(s_k)\right\}dF_\bX(\bx)
\end{align*}
and the result in the text follows by noting that for any $j = 1, ..., t$,
\begin{align*}
    \Qsc_{gj}^*(\bX, \bSbar_{j-1}) &= \int \prod_{k=j}^t\mu_{gk}(\bX, \bSbar_{k-1})d\left\{\prod_{k=j}^{t_0} F_{S_k | T > k}(s_k)\right\}
\end{align*}

The IPW-type identification also follows from a similar argument as the one for $\Deltat$. Recall that the claim is that \begin{align*}
    \Deltastt &= \E\left\{ 
                    \frac{G}{e(\bX)} \frac{A_{i1}Y_{i1}}{\gamma_{11}(\bX)}\prod_{k=2}^{t_0+1} \frac{A_{ik}Y_{ik}\pi^*_{k-1}(\bX, \bSbar_{k-2})}{\gamma_{1k}(\bX, \bSbar_{k-1})\pi_{k-1}(\bX, \bSbar_{k-1})}\prod_{k=t_0+2}^t \frac{A_{ik}Y_{ik}}{\gamma_{1k}(\bX, \bSbar_{k-1})} -\right.\\
                    &\qquad\left.\frac{1-G}{1-e(\bX)}\frac{A_{ik}Y_{ik}}{\gamma_{0k}(\bX, \bSbar_{k-1})} \prod_{k=2}^{t_0+1} \frac{A_{ik}Y_{ik}\{1-\pi^*_{k-1}(\bX, \bSbar_{k-2})\}}{\gamma_{0k}(\bX, \bSbar_{k-1})\{1-\pi_{k-1}(\bX, \bSbar_{k-1})\}}
                    \prod_{k=t_0+2}^t \frac{A_{ik}Y_{ik}}{\gamma_{0k}(\bX, \bSbar_{k-1})}
                  \right\}\\
                  &= \Psi_1 - \Psi_0.
\end{align*}
Taking the first term first, we can rewrite it as
\begin{align*}
    \Psi_1 &= \int
                    \frac{g}{e(\bx)} \frac{a_1y_1}{\gamma_{11}(\bx)}\prod_{k=2}^{t_0+1} \frac{a_ky_k\pi^*_{k-1}(\bx, \bsbar_{k-2})}{\gamma_{1k}(\bx, \bsbar_{k-1})\pi_{k-1}(\bx, \bsbar_{k-1})}
                  \prod_{k=t_0+2}^t \frac{a_ky_k}{\gamma_{1k}(\bx, \bsbar_{k-1})}\times\\
                  &\qquad 
                  f_{\bX G \bAbar_t\bYbar_t\bSbar_{t-1}}(\bx, g, \babar_t, \bybar_t, \bsbar_{t-1})d\bybar_td\babar_td\bsbar_{t-1}dgd\bx\\
                  &= \int
                    gf_\bX(\bx) \frac{a_1y_1}{f_{\bX G A_1}(\bx, 1, 1)}\times\\
                    &\qquad 
                    \prod_{k=2}^{t_0+1} 
                        \frac{a_ky_k
                            f_{\bX G\bAbar_{k-1}\bYbar_{k-1}\bSbar_{k-2}}(\bx, 1, \bone_{k-1}, \bone_{k-1}, \bsbar_{k-2})
                            f_{\bX\bAbar_{k-1}\bYbar_{k-1}\bSbar_{k-1}}(\bx, \bone_{k-1}, \bone_{k-1}, \bsbar_{k-1}) 
                            }{
                            f_{\bX G\bAbar_k\bYbar_{k-1}\bSbar_{k-1}}(\bx, 1, \bone_k, \bone_{k-1}, \bsbar_{k-1})
                            f_{\bX\bAbar_{k-1}\bYbar_{k-1}\bSbar_{k-2}}(\bx, \bone_{k-1}, \bone_{k-1}, \bsbar_{k-2})}
                  \times\\
                  &\qquad \prod_{k=t_0+2}^{t} \frac{a_ky_k
                            f_{\bX G\bAbar_{k-1}\bYbar_{k-1}\bSbar_{t_0}}(\bx, 1, \bone_{k-1}, \bone_{k-1}, \bsbar_{t_0})
                            }{
                            f_{\bX G\bAbar_k\bYbar_{k-1}\bSbar_{t_0}}(\bx, 1, \bone_k, \bone_{k-1}, \bsbar_{t_0})
                            }\times\\
                  &\qquad\qquad f_{\bX G\bAbar_t\bYbar_t\bSbar_{t_0}}(\bx, g, \babar_t, \bybar_t, \bsbar_{t_0})d\bybar_td\babar_td\bsbar_{t_0}dgd\bx\\
                  &= \int gf_\bX(\bx) \prod_{k=1}^{t_0} a_ky_k \frac{f_{\bX G\bAbar_k\bYbar_k\bSbar_{k-1}}(\bx, 1, \bone_{k}, \bone_{k}, \bsbar_{k-1})}{f_{\bX G\bAbar_k\bYbar_{k-1}\bSbar_{k-1}}(\bx, 1, \bone_k, \bone_{k-1}, \bsbar_{k-1})}\times
                  \frac{f_{\bX\bAbar_k\bYbar_k\bSbar_k}(\bx, \bone_k, \bone_k, \bsbar_k)}{f_{\bX\bAbar_k\bYbar_k\bSbar_{k-1}}(\bx, \bone_k, \bone_k, \bsbar_{k-1})}\times\\
                  &\qquad \prod_{k=t_0+1}^{t-1} a_ky_k \frac{f_{\bX G\bAbar_k\bYbar_k\bSbar_{t_0}}(\bx, 1, \bone_{k}, \bone_{k}, \bsbar_{t_0})}{f_{\bX G\bAbar_k\bYbar_{k-1}\bSbar_{t_0}}(\bx, 1, \bone_k, \bone_{k-1}, \bsbar_{t_0})}\times a_t y_t\frac{f_{\bX G\bAbar_t\bYbar_t\bSbar_{t_0}}(\bx, 1, \babar_t, \bybar_{t}, \bsbar_{t_0})}{f_{\bX G\bAbar_t\bYbar_{t-1}\bSbar_{t_0}}(\bx, 1, \bone_t, \bone_{t-1}, \bsbar_{t_0})}d\bybar_td\babar_td\bsbar_{t_0}dgd\bx\\
                  &= \E_\bX\left[\int \prod_{k=1}^t \mu_{1k}(\bX, \bsbar_{k-1})f_{\bSbar_{k-1} | \bAbar_{k-1} = \bYbar_{k-1} = 1, \bX}(\bsbar_{k-1}, \bX) d\bsbar_{t_0}\right]
\end{align*}
\subsection{Influence function for $\Deltat$ \label{if_delta_second}}
The components of the influence function can be computed by finding $d \Psi_1 / d f$ for $f = f_\bX, f_{\bX G}, f_{\bX G\bAbar_{k-1}\bYbar_{k-1}\bSbar_{k-1}}, f_{\bX G\bAbar_k\bYbar_{k-1}\bSbar_{k-1}},$ and $f_{\bX G\bAbar_t\bYbar_t\bSbar_{t-1}}$. 
\begin{align*}
    \frac{d}{d f_{\bX}} \Psi_1 &= \int\left\{ 
                    \frac{g}{f_{\bX G}(\bX, 1)} \prod_{k=1}^t \frac{a_ky_kf_{\bX G\bAbar_{k-1}\bYbar_{k-1}\bSbar_{k-1}}(\bX, 1, \bone_{k-1}, \bone_{k-1}, \bsbar_{k-1}) }{f_{\bX G\bAbar_k\bYbar_{k-1}\bSbar_{k-1}}(\bX, 1, \bone_{k}, \bone_{k-1} \bsbar_{k-1})}
                  \right\}\times\\
                  &\qquad\qquad f_{\bX G\bAbar_t\bYbar_t\bSbar_{t-1}}(\bX, g, \babar_t, \bybar_t, \bsbar_{t-1})d\bybar_td\babar_td\bsbar_{t-1}dg\\
                  &= \int g \times a_1y_1 \frac{f_{\bX G A_1 Y_1 S_1}(\bX, 1, 1, 1, s_1)}{f_{\bX GA_1}(\bX, 1, 1)} \times a_2y_2 \frac{f_{\bX G \bAbar_2 \bYbar_2 \bSbar_2}(\bX, 1, \bone_2, \bone_2, \bsbar_2)}{f_{\bX G\bAbar_2Y_1S_1}(\bX, 1, \bone_2, 1, s_1)} \times ... \\ & \qquad \times a_ty_t \frac{f_{\bX G \bAbar_t \bYbar_t \bSbar_{t-1}}(\bX, 1, \bone_t, \bone_t, \bsbar_{t-1})}{f_{\bX G\bAbar_tY_{t-1}S_{t-1}}(\bX, 1, \bone_t, \bone_{t-1}, \bsbar_{t-1})}d\bybar_td\babar_td\bsbar_{t-1}dg\\
                  &= \mu_{11}(\bX)\Qsc_{11}(\bX).
\end{align*}
\begin{align*}
    \frac{d}{d f_{\bX G}} \Psi_1 &= -\int\left\{ 
                    \frac{G}{e(\bX)f_{\bX G}(\bX, 1)}\prod_{k=1}^t \frac{a_ky_kf_{\bX G\bAbar_{k-1}\bYbar_{k-1}\bSbar_{k-1}}(\bX, 1, \bone_{k-1}, \bone_{k-1}, \bsbar_{k-1}) }{f_{\bX G\bAbar_k\bYbar_{k-1}\bSbar_{k-1}}(\bX, 1, \bone_{k}, \bone_{k-1} \bsbar_{k-1})}
                  \right\}\times\\
                  &\qquad\qquad f_{\bX G\bAbar_t\bYbar_t\bSbar_{t-1}}(\bX, G, \babar_t, \bybar_t, \bsbar_{t-1})d\bybar_td\babar_td\bsbar_{t-1}dg\\
                  &= -\frac{G}{e(\bX)}\mu_{11}(\bX)\Qsc_{11}(\bX).
\end{align*}
For $j = 1, ..., t$ we have
\begin{align*}
    \frac{d}{d f_{\bX G\bAbar_{j-1}\bYbar_{j-1}\bSbar_{j-1}}} \Psi_1 &= \int\left\{ 
                    \frac{G}{e(\bX)}\prod_{k=1}^{j-1} \frac{A_kY_kf_{\bX G\bAbar_{k-1}\bYbar_{k-1}\bSbar_{k-1}}(\bX, 1, \bone_{k-1}, \bone_{k-1}, \bSbar_{k-1}) }{f_{\bX G\bAbar_k\bYbar_{k-1}\bSbar_{k-1}}(\bX, 1, \bone_{k}, \bone_{k-1}, \bSbar_{k-1})}
                  \right\}\times\\
                  &\frac{a_jy_j}{f_{\bX G\bAbar_j\bYbar_{j-1}\bSbar_{j-1}}(\bX, 1, \bone_{j}, \bone_{j-1}, \bSbar_{j-1})}\\
                  &\prod_{k=j+1}^{t} \frac{a_ky_kf_{\bX G\bAbar_{k-1}\bYbar_{k-1}\bSbar_{k-1}}(\bX, 1, \bone_{k-1}, \bone_{k-1}, (\bSbar_{j-1}, \bsbar_j^{k-1})) }{f_{\bX G\bAbar_k\bYbar_{k-1}\bSbar_{k-1}}(\bX, 1, \bone_{k}, \bone_{k-1}, \bSbar_{k-1})}\times\\
                  &\qquad\qquad f_{\bX G\bAbar_t\bYbar_t\bSbar_{t-1}}(\bX, G, \babar_t, \bybar_t, \bsbar_{t-1})d\bybar_td\babar_td\bsbar_{t-1}dg\\
                  &= \frac{G}{e(\bX)}\prod_{k=1}^{j-1} \frac{A_kY_k}{\gamma_{1k}(\bX, \bSbar_{k-1})}\mu_{1j}(\bX, \bSbar_{j-1})\Qsc_{1j}(\bX, \bSbar_{j-1})
\end{align*}
\begin{align*}
    \frac{d}{d f_{\bX G\bAbar_j\bYbar_{j-1}\bSbar_{j-1}}} \Psi_1 &= -\int\left\{ 
                    \frac{G}{e(\bX)}\prod_{k=1}^{j-1} \frac{A_kY_kf_{\bX G\bAbar_{k-1}\bYbar_{k-1}\bSbar_{k-1}}(\bX, 1, \bone_{k-1}, \bone_{k-1}, \bSbar_{k-1}) }{f_{\bX G\bAbar_k\bYbar_{k-1}\bSbar_{k-1}}(\bX, 1, \bone_{k}, \bone_{k-1}, \bSbar_{k-1})}
                  \right\}\times\\
                  &\frac{a_jy_jf_{\bX G\bAbar_{j-1}\bYbar_{j-1}\bSbar_{j-1}}(\bX, 1, \bone_{j}, \bone_{j-1}, \bSbar_{j-1})}{f^2_{\bX G\bAbar_j\bYbar_{j-1}\bSbar_{j-1}}(\bX, 1, \bone_{j}, \bone_{j-1}, \bSbar_{j-1})}\\
                  &\prod_{k=j+1}^{t} \frac{a_ky_kf_{\bX G\bAbar_{k-1}\bYbar_{k-1}\bSbar_{k-1}}(\bX, 1, \bone_{k-1}, \bone_{k-1}, (\bSbar_{j-1}, \bsbar_j^{k-1})) }{f_{\bX G\bAbar_k\bYbar_{k-1}\bSbar_{k-1}}(\bX, 1, \bone_{k}, \bone_{k-1}, \bSbar_{k-1})}\times\\
                  &\qquad\qquad f_{\bX G\bAbar_t\bYbar_t\bSbar_{t-1}}(\bX, G, \babar_t, \bybar_t, \bsbar_{t-1})d\bybar_td\babar_td\bsbar_{t-1}dg\\
                  &= -\frac{G}{e(\bX)}\prod_{k=1}^{j-1} \frac{A_kY_k}{\gamma_{1k}(\bX, \bSbar_{k-1})}\frac{A_j\mu_{1j}(\bX, \bSbar_{j-1})\Qsc_{1j}(\bX, \bSbar_{j-1})}{\gamma_{1j}(\bX, \bSbar_{j-1})}
\end{align*}
Noting that 
\begin{align*}
    \frac{d}{d f_{\bX G\bAbar_t\bYbar_t\bSbar_{t-1}}} \Psi_1 = \frac{G}{e(\bX)} \prod_{k=1}^t \frac{A_{ik}Y_{ik}}{\gamma_{1k}(\bX, \bSbar_{k-1})}
\end{align*}
along with corresponding computations on $\Psi_0$ yields the form of the influence function shown in Section \ref{sec:main-setting}.

\subsection{Influence function for $\Deltastt$ \label{if_deltas_second}}

As noted in the main text, 
\begin{align*}
    \Deltastt &= \E\left\{ 
                    \frac{G}{e(\bX)} \frac{A_{i1}Y_{i1}}{\gamma_{11}(\bX)}\prod_{k=2}^{t_0+1} \frac{A_{ik}Y_{ik}\pi^*_{k-1}(\bX, \bSbar_{k-2})}{\gamma_{1k}(\bX, \bSbar_{k-1})\pi_{k-1}(\bX, \bSbar_{k-1})}\prod_{k=t_0+2}^t \frac{A_{ik}Y_{ik}}{\gamma_{1k}(\bX, \bSbar_{k-1})} -\right.\\
                    &\qquad\left.\frac{1-G}{1-e(\bX)}\frac{A_{ik}Y_{ik}}{\gamma_{0k}(\bX, \bSbar_{k-1})} \prod_{k=2}^{t_0+1} \frac{A_{ik}Y_{ik}\{1-\pi^*_{k-1}(\bX, \bSbar_{k-2})\}}{\gamma_{0k}(\bX, \bSbar_{k-1})\{1-\pi_{k-1}(\bX, \bSbar_{k-1})\}}
                    \prod_{k=t_0+2}^t \frac{A_{ik}Y_{ik}}{\gamma_{0k}(\bX, \bSbar_{k-1})}
                  \right\}\\
                  &= \Psi_1 - \Psi_0.
\end{align*}
Taking the first term first, we can rewrite it as
\begin{align*}
    \Psi_1 &= \int
                    \frac{g}{e(\bx)} \frac{a_1y_1}{\gamma_{11}(\bx)}\prod_{k=2}^{t_0+1} \frac{a_ky_k\pi^*_{k-1}(\bx, \bsbar_{k-2})}{\gamma_{1k}(\bx, \bsbar_{k-1})\pi_{k-1}(\bx, \bsbar_{k-1})}
                  \prod_{k=t_0+2}^t \frac{a_ky_k}{\gamma_{1k}(\bx, \bsbar_{k-1})}\times\\
                  &\qquad 
                  f_{\bX G \bAbar_t\bYbar_t\bSbar_{t-1}}(\bx, g, \babar_t, \bybar_t, \bsbar_{t-1})d\bybar_td\babar_td\bsbar_{t-1}dgd\bx\\
                  &= \int
                    gf_\bX(\bx) \frac{a_1y_1}{f_{\bX G A_1}(\bx, 1, 1)}\times\\
                    &\qquad 
                    \prod_{k=2}^{t_0+1} 
                        \frac{a_ky_k
                            f_{\bX G\bAbar_{k-1}\bYbar_{k-1}\bSbar_{k-2}}(\bx, 1, \bone_{k-1}, \bone_{k-1}, \bsbar_{k-2})
                            f_{\bX\bAbar_{k-1}\bYbar_{k-1}\bSbar_{k-1}}(\bx, \bone_{k-1}, \bone_{k-1}, \bsbar_{k-1}) 
                            }{
                            f_{\bX G\bAbar_k\bYbar_{k-1}\bSbar_{k-1}}(\bx, 1, \bone_k, \bone_{k-1}, \bsbar_{k-1})
                            f_{\bX\bAbar_{k-1}\bYbar_{k-1}\bSbar_{k-2}}(\bx, \bone_{k-1}, \bone_{k-1}, \bsbar_{k-2})}
                  \times\\
                  &\qquad \prod_{k=t_0+2}^{t} \frac{a_ky_k
                            f_{\bX G\bAbar_{k-1}\bYbar_{k-1}\bSbar_{k-1}}(\bx, 1, \bone_{k-1}, \bone_{k-1}, \bsbar_{k-1})
                            }{
                            f_{\bX G\bAbar_k\bYbar_{k-1}\bSbar_{k-1}}(\bx, 1, \bone_k, \bone_{k-1}, \bsbar_{k-1})
                            }\times\\
                  &\qquad\qquad f_{\bX G\bAbar_t\bYbar_t\bSbar_{t-1}}(\bx, g, \babar_t, \bybar_t, \bsbar_{t-1})d\bybar_td\babar_td\bsbar_{t-1}dgd\bx\\
                  &= \int gf_\bX(\bx) \prod_{k=1}^{t_0} a_ky_k \frac{f_{\bX G\bAbar_k\bYbar_k\bSbar_{k-1}}(\bx, 1, \bone_{k}, \bone_{k}, \bsbar_{k-1})}{f_{\bX G\bAbar_k\bYbar_{k-1}\bSbar_{k-1}}(\bx, 1, \bone_k, \bone_{k-1}, \bsbar_{k-1})}\times
                  \frac{f_{\bX\bAbar_k\bYbar_k\bSbar_k}(\bx, \bone_k, \bone_k, \bsbar_k)}{f_{\bX\bAbar_k\bYbar_k\bSbar_{k-1}}(\bx, \bone_k, \bone_k, \bsbar_{k-1})}\times\\
                  &\qquad \prod_{k=t_0+1}^{t-1} a_ky_k \frac{f_{\bX G\bAbar_k\bYbar_k\bSbar_{k-1}}(\bx, 1, \bone_{k}, \bone_{k}, \bsbar_{k-1})}{f_{\bX G\bAbar_k\bYbar_{k-1}\bSbar_{k-1}}(\bx, 1, \bone_k, \bone_{k-1}, \bsbar_{k-1})}\times
                  \frac{f_{\bX G\bAbar_k\bYbar_k\bSbar_k}(\bx, 1, \bone_k, \bone_k, \bsbar_k)}{f_{\bX G, \bAbar_k\bYbar_k\bSbar_{k-1}}(\bx,1, \bone_k, \bone_k, \bsbar_{k-1})}\times\\
                  &\qquad\qquad a_t y_t\frac{f_{\bX G\bAbar_t\bYbar_t\bSbar_{t-1}}(\bx, 1, \babar_t, \bybar_{t}, \bsbar_{t-1})}{f_{\bX G\bAbar_t\bYbar_{t-1}\bSbar_{t-1}}(\bx, 1, \bone_t, \bone_{t-1}, \bsbar_{t-1})}d\bybar_td\babar_td\bsbar_{t-1}dgd\bx.
\end{align*}

Following similar arguments as for the influence function of $\Deltat$ in the previous section, we have that 
\begin{align*}
    \frac{d}{d f_{\bX G\bAbar_t\bYbar_t\bSbar_{t-1}}} \Psi_1 &= \frac{G}{e(\bX)} \prod_{k=1}^{t_0+1} \frac{A_{k}Y_{k}\pi^*_{k-1}(\bX, \bSbar_{k-2})}{\gamma_{1k}(\bX, \bSbar_{k-1})\pi_{k-1}(\bX, \bSbar_{k-1})}\prod_{k=t_0+2}^t \frac{A_{k}Y_{k}}{\gamma_{1k}(\bX, \bSbar_{k-1})}.
\end{align*}
And for $j = t_0+1, ..., t$
\begin{align*}
    \frac{d}{d f_{\bX G\bAbar_j\bYbar_{j-1}\bSbar_{j-1}}} \Psi_1 &= -\frac{G}{e(\bX)} \prod_{k=1}^{t_0+1} \frac{A_{k}Y_{k}\pi^*_{k-1}(\bX, \bSbar_{k-2})}{\gamma_{1k}(\bX, \bSbar_{k-1})\pi_{k-1}(\bX, \bSbar_{k-1})}\prod_{k=t_0+2}^{j-1} \frac{A_{k}Y_{k}}{\gamma_{1k}(\bX, \bSbar_{k-1})}\frac{A_j\mu_{1j}(\bX, \bSbar_{j-1})\Qsc^*_{1j}(\bX, \bSbar_{j-1})}{\gamma_{1j}(\bX, \bSbar_{j-1})}
\end{align*} and for $j = t_0 + 2, ..., t$
\begin{align*}
    \frac{d}{d f_{\bX G\bAbar_{j-1}\bYbar_{j-1}\bSbar_{j-1}}} \Psi_1 &=  \frac{G}{e(\bX)} \prod_{k=1}^{t_0+1} \frac{A_{k}Y_{k}\pi^*_{k-1}(\bX, \bSbar_{k-2})}{\gamma_{1k}(\bX, \bSbar_{k-1})\pi_{k-1}(\bX, \bSbar_{k-1})}\prod_{k=t_0+2}^{j-1} \frac{A_{k}Y_{k}}{\gamma_{1k}(\bX, \bSbar_{k-1})}\mu_{1j}(\bX, \bSbar_{j-1})\Qsc^*_{1j}(\bX, \bSbar_{j-1}).
\end{align*}
For $j = 1, ..., t_0$, 
\begin{align*}
    \frac{d}{d f_{\bX \bAbar_{j}\bYbar_{j}\bSbar_{j}}} \Psi_1 &=  \frac{1}{e(\bX)} \prod_{k=1}^{j} \frac{A_{k}Y_{k}\pi^*_{k-1}(\bX, \bSbar_{k-2})}{\gamma_{1k}(\bX, \bSbar_{k-1})\pi_{k-1}(\bX, \bSbar_{k-1})}\pi^*_{j}(\bX, \bSbar_{j-1})\mu_{1j+1}(\bX, \bSbar_{j})\Qsc^*_{1j+1}(\bX, \bSbar_{j}),
\end{align*}
\begin{align*}
    \frac{d}{d f_{\bX G\bAbar_j\bYbar_{j}\bSbar_{j-1}}} \Psi_1 &= \frac{G}{e(\bX)} \prod_{k=1}^{j} \frac{A_{k}Y_{k}\pi^*_{k-1}(\bX, \bSbar_{k-2})}{\gamma_{1k}(\bX, \bSbar_{k-1})\pi_{k-1}(\bX, \bSbar_{k-1})}\Qsc^*_{1j}(\bX, \bSbar_{j-1}),
\end{align*}
\begin{align*}
    \frac{d}{d f_{\bX\bAbar_j\bYbar_{j}\bSbar_{j-1}}} \Psi_1 &= -\frac{1}{e(\bX)} \prod_{k=1}^{j} \frac{A_{k}Y_{k}\pi^*_{k-1}(\bX, \bSbar_{k-2})}{\gamma_{1k}(\bX, \bSbar_{k-1})\pi_{k-1}(\bX, \bSbar_{k-1})}\pi^*_j(\bX, \bSbar_{j-1})\Qsc^*_{1j}(\bX, \bSbar_{j-1})
\end{align*}
\begin{align*}
    \frac{d}{d f_{\bX G\bAbar_j\bYbar_{j-1}\bSbar_{j-1}}} \Psi_1 &= -\frac{G}{e(\bX)} \prod_{k=1}^{j-1} \frac{A_{k}Y_{k}\pi^*_{k-1}(\bX, \bSbar_{k-2})}{\gamma_{1k}(\bX, \bSbar_{k-1})\pi_{k-1}(\bX, \bSbar_{k-1})}\frac{A_{j}\pi^*_{j-1}(\bX, \bSbar_{j-2})\mu_{1j}(\bX, \bSbar_{j-1})\Qsc^*_{1j}(\bX, \bSbar_{j-1})}{\gamma_{1j}(\bX, \bSbar_{j-1})\pi_{j-1}(\bX, \bSbar_{j-1})},
\end{align*}
A symmetric argument for $\Psi_0$ yields the form of the influence function in the main text.

\section{Targeted minimum loss-based estimation \label{app:tmle}}

\subsection{TML estimation for $\Deltat$}
The TML estimator for $\Deltat$ proceeds as follows. First, rewrite the influence as
\begin{align*}
    \phi\{O, \Deltat, \Psi\} = \sum_{g=0}^1 (-1)^{1-g} \left\{\sum_{j=1}^tH_{gj} (Q_{Y gj} - Q_{\mu gj}) + Q_{\mu g1}\right\} - \Deltat
\end{align*}
where $H_{gj} = \frac{I\{G = g\}}{e_g}\prod_{k=1}^{j-1}\frac{A_k Y_k}{\gamma_{1k}}\frac{A_j}{\gamma_{gj}}, e_g = e^g(1-e)^{1-g}, Q_{Ygj} = Y_j\mu_{gj+1}\Qsc_{gj+1}$ and $Q_{\mu gj} = \mu_{gj}\Qsc_{gj}.$

We first obtain estimates of $\Psi$ using cross-fitting. we next seek to obtain targeted versions of $Q_{\mu gk}, j = , 1, ..., t; g = 0, 1.$ We initialize $Q_{\mu gt+1} = 1$, and we assume that for $k = t-1, ..., 1$, we have already obtained a targeted version of $Q_{\mu gk+1}$, called $\Qhat_{\mu gk+1}$
\begin{enumerate}
    \item Compute $Q_{Y gk} = Y_k\Qhat_{\mu gk+1}$.
    \item Regress $Q_{Y gk}$ on $\bX, \bSbar_{k-1}$ among those still at risk in  treatment group $g$ at time $k$.
    \item Obtain predictions for all individuals still at risk at time $k$ in both treatment arms to obtain $\Qtilde_{\mu gk}$, an initial estimate of $Q_{\mu gk}$.
    \item Update this estimate using the estimated intercept $\epsilon$ from an intercept-only weighted logistic regression of $Q_{Y gk}$ with offset $\text{logit}(\Qtilde_{\mu gk})$ and weights $H_{gk}$. The updated estimate is $\Qhat_{\mu gk} = \text{expit} \left\{ \text{logit}(\Qtilde_{\mu gk}) + \epsilon \right\}$.
\end{enumerate}
Finally, the TML estimator of $\Deltat$ is
\begin{align*}
    \Deltahat(t) = n\inv\sumin \left\{\Qhat_{\mu 11}(\bX_i) - \Qhat_{\mu 01}(\bX_i)\right\}
\end{align*}

\subsection{TML estimator for $\Deltastt$}
The TML estimation for $\Deltastt$ follows the same broad outlines of the estimation for $\Deltat$ but must has an additional step. First, we rewrite the influence function as 
\begin{align*}
    \phi_S\{O, \Deltastt, \Psi\} = \sum_{g=0}^1 (-1)^{1-g} \left[\left\{\sum_{j=1}^t H^*_{Ygj} (Q^*_{Ygj} - Q^*_{\mu gj}) + \sum_{j=1}^{t_0}H^*_{Sgj} (Q^*_{\mu gj+1} - \Qsc^*_{gj})\right\} + Q^*_{\mu g1}\right] - \Deltat
\end{align*}
where 
\begin{align*}
    H_{Y gj}^* &= \frac{I\{G = g\}}{e_g} \prod_{k=1}^{j-1} \frac{A_k Y_k \pi^*_{gk-1}}{\gamma_{gk}\pi_{gk-1}} \left[\frac{\pi^*_{gj-1}A_j}{\pi_{gj-1}\gamma_{gj}} \right],\\
    Q^*_{Y gj} &= Y_j\Qsc^*_{gj}, \\
    Q^*_{\mu gj} &= \mu_j\Qsc^*_{gj}, \\
    H^*_{S gj} &= e_g\inv \prod_{k=1}^{j}\frac{\pi^*_{gk-1}A_kY_k}{\pi_{gk-1}\gamma_{gk}}\pi^*_{gj-1}, \\
    \pi_{gk} &= \pi_k^g(1-\pi_k)^{1-g}, \pi^*_{gk} = \pi_k^{*g}(1-\pi^*_k)^{1-g}
\end{align*}.

We similarly want to obtain targeted versions of $Q^*_{\mu gj}$, which we obtain with the following algorithm. We first obtain estimates of $\Psi$ using cross-fitting, as above. To obtain targeted versions of $Q^*_{\mu gk}, j = , 1, ..., t; g = 0, 1,$ we initialize $Q^*_{\mu gt+1} = 1$, and we assume that for $k = t-1, ..., 1$, we have already obtained a targeted version of $Q^*_{\mu gk+1}$, which we call $\Qhat^*_{\mu gk+1}$.
The targeting proceeds differently for $k > t_0$ (when there is no surrogate information) and for $k \leq t_0$ (when there is surrogate information). For $k > t_0,$
\begin{enumerate}
    \item Compute $Q^*_{Y gk} = Y_k\Qhat^*_{\mu gk+1}$.
    \item Regress $Q^*_{Y gk}$ on $\bX, \bSbar_{t_0}$ among those still at risk in treatment group $g$ at time $k$.
    \item Obtain predictions for all individuals still at risk at time $k$ in both treatment arms to obtain $\Qtilde^*_{\mu gk}$, an initial estimate of $Q^*_{\mu gk}$.
    \item Update this estimate using the estimated intercept $\epsilon$ from an intercept-only weighted logistic regression of $Q^*_{Y gk}$ with offset $\text{logit}(\Qtilde^*_{\mu gk})$ and weights $H^*_{Ygk}$. The updated estimate is $\Qhat^*_{\mu gk} = \text{expit} \left\{ \text{logit}(\Qtilde^*_{\mu gk}) + \epsilon \right\}$.
\end{enumerate}
For $k \leq t_0$, the algorithm proceeds in two stages, by first targeting $\Qsc^*_{gk}$ in both treatment groups, then targeting $Q^*_{\mu gk}$:
\begin{enumerate}
    \item Regress $\Qhat^*_{\mu gk+1}$ on $\bX, \bSbar_{k-1}$ among those still at risk in both treatment groups at time $k$. 
    \item Obtain predictions for all individuals still at risk at time $k$ in both treatment arms to obtain $\Qsctilde^*_{gk}$, an initial estimate of $\Qsc^*_{gk}$.
    \item Update this estimate using the estimated intercept $\epsilon_1$ from an intercept-only weighted logistic regression of $\Qhat^*_{\mu gk+1}$ with offset $\text{logit}(\Qsctilde^*_{gk})$ and weights $H^*_{Sgk}$. The updated estimate is $\Qschat^*_{gk} = \text{expit} \left\{ \text{logit}(\Qsctilde^*_{gk}) + \epsilon_1 \right\}$.
    \item Compute $Q^*_{Y gk} = Y_k\Qschat^*_{gk}$.
    \item Regress $Q^*_{Y gk}$ on $\bX, \bSbar_{k-1}$ among those still at risk in treatment group $g$ at time $k$.
    \item Obtain predictions for all individuals still at risk at time $k$ in both treatment arms to obtain $\Qtilde^*_{\mu gk}$, an initial estimate of $Q^*_{\mu gk}$.
    \item Update this estimate using the estimated intercept $\epsilon_2$ from an intercept-only weighted logistic regression of $Q^*_{Y gk}$ with offset $\text{logit}(\Qtilde^*_{\mu gk})$ and weights $H^*_{Ygk}$. The updated estimate is $\Qhat^*_{\mu gk} = \text{expit} \left\{ \text{logit}(\Qtilde^*_{\mu gk}) + \epsilon_2 \right\}$.
\end{enumerate}
Finally, the TML estimator of $\Deltastt$ is
\begin{align*}
    \Deltahat_S(t, t_0) = n\inv\sumin \left\{\Qhat^*_{\mu 11}(\bX_i) - \Qhat^*_{\mu 01}(\bX_i)\right\}
\end{align*}

\section{Asymptotic normality and inference}\label{app:asymptotics}
\subsection{Double robustness}
We demonstrate the results in the setting with $t = 3$ and $t_0 = 2$ so that the argument is clear while maintaining all relevant complexity. Extension to $t > 3, t_0 > 2$ is straightforward.

 Write a version of the influence function with estimated nuisance functions $\Psihat$:
\begin{align*}
    \phihat &= \phi(\bO, \Deltat, \Psihat) = \frac{G}{\ehat(\bX)}\left[ 
        \frac{A_1}{\gammahat_{11}(\bX)}\{Y_1\muhat_{12}(\bX, S_1)\Qschat_{12}(\bX, S_1) - \muhat_{11}(\bX)\Qschat_{11}(\bX)\} +\right.\\
        &\frac{A_1Y_1A_2}{\gammahat_{11}(\bX)\gammahat_{12}(\bX, S_1)}\{Y_2\muhat_{13}(\bX, \bSbar_2) - \muhat_{12}(\bX, S_1)\Qschat_{12}(\bX, S_1) \} +\\
        &\left.\frac{A_1Y_1A_2Y_2A_3}{\gammahat_{11}(\bX)\gammahat_{12}(\bX, S_1)\gammahat_{13}(\bX, \bSbar_2)}\{Y_3 - \muhat_{13}(\bX, \bSbar_2)\} 
    \right] - \\
    &\frac{1-G}{1-\ehat(\bX)}\left[ 
        \frac{A_1}{\gammahat_{01}(\bX)}\{Y_1\muhat_{02}(\bX, S_1)\Qschat_{02}(\bX, S_1) - \muhat_{01}(\bX)\Qschat_{01}(\bX)\} +\right.\\
        &\frac{A_1Y_1A_2}{\gammahat_{01}(\bX)\gammahat_{02}(\bX, S_1)}\{Y_2\muhat_{03}(\bX, \bSbar_2) - \muhat_{02}(\bX, S_1)\Qschat_{02}(\bX, S_1) \} +\\
        &\left.\frac{A_1Y_1A_2Y_2A_3}{\gammahat_{01}(\bX)\gammahat_{02}(\bX, S_1)\gammahat_{03}(\bX, \bSbar_2)}\{Y_3 - \muhat_{03}(\bX, \bSbar_2)\} 
    \right] + \muhat_{11}(\bX)\Qschat_{11}(\bX) - \muhat_{01}(\bX)\Qschat_{01}(\bX) - \Deltat\\
    &= \varphi_1 - \varphi_0 + \muhat_{11}(\bX)\Qschat_{11}(\bX) - \muhat_{01}(\bX)\Qschat_{01}(\bX) - \Deltat.
\end{align*}
Note that we use sample splitting to estimate the nuisance functions and throughout the rest of the text use a slightly more complex notation to indicate this, but how the nuisance functions are estimated is not material to this proof.

Define the events $\Usc_1, \Usc_2(S_1)$ such that for any random variable $B$, we have $\E(B | \Usc_1) = \E(B | \bX, G = 1, A_1 = Y_1 = 1), \E\{B | \Usc_2(S_1)\} = \E(B |\bX, G = 1, A_1 = Y_1 = A_2, Y_2 = 1, S_1)$. Looking at the expected value of $\varphi_1$, we have:
\begin{align*}
    \E(\varphi_1) &= \E\left(\frac{e(\bX)}{\ehat(\bX)}\left[ 
        \frac{\gamma_{11}(\bX)}{\gammahat_{11}(\bX)}\{\mu_{11}(\bX)\E\left\{\muhat_{12}(\bX, S_1)\Qschat_{12}(\bX, S_1) | \Usc_1\right\} - \muhat_{11}(\bX)\Qschat_{11}(\bX)\} +\right.\right.\\
        &\qquad \frac{\gamma_{11}(\bX)\mu_{11}(\bX)}{\gammahat_{11}(\bX)}\E\left(\frac{\gamma_{12}(\bX, S_1)}{\gammahat_{12}(\bX, S_1)}\left[\mu_{12}(\bX, S_1)\E\left\{\muhat_{13}(\bX, \bSbar_2)| \Usc_2(S_1)\right\}   - \muhat_{12}(\bX, S_1)\Qschat_{12}(\bX, S_1) \right] | \Usc_1\right) +\\
        &\qquad\left.\left.\frac{\gamma_{11}(\bX)\mu_{11}(\bX)}{\gammahat_{11}(\bX)}\E\left(\frac{\gamma_{12}(\bX, S_1)\mu_{12}(\bX, S_1)}{\gammahat_{12}(\bX, S_1)}\E\left\{\gamma_{13}(\bX, \bSbar_2)\frac{\mu_{13}(\bX, \bSbar_2) - \muhat_{13}(\bX, \bSbar_2)}{\gammahat_{13}(\bX, \bSbar_2)} | \Usc_2(S_1)\right\} | \Usc_1\right) 
    \right]\right)
\end{align*}
If $\muhat_{ak} = \mu_{ak}$ and $\Qschat_{ak} = \Qsc_{ak}$, then it is easy to see from the above that
\begin{align*}
    \E(\varphi_1) = 0,
\end{align*}
and a similar argument holds for $\E(\varphi_0)$, which suggests that if $\muhat_{ak} = \mu_{ak}$ and $\Qschat_{ak} = \Qsc_{ak}$, we have 
\begin{align*}
    \E(\phihat) = \E\left\{\mu_{11}(\bX)\Qsc_{11}(\bX) - \mu_{01}(\bX)\Qsc_{01}(\bX)\right\} - \Deltat = 0.
\end{align*}

On the other hand, if $\ehat(\bx) = e(\bx)$ and $\gammahat_{ak} = \gamma_{ak}$, we have
\begin{align*}
    \E(\varphi_1) &= \E\left(\mu_{11}(\bX)\E\left\{\muhat_{12}(\bX, S_1)\Qschat_{12}(\bX, S_1) | \Usc_1\right\} - \muhat_{11}(\bX)\Qschat_{11}(\bX)\right) + \\
        &\qquad \E\left(\mu_{11}(\bX)\E\left[\{\mu_{12}(\bX, S_1)\E\left\{\muhat_{13}(\bX, \bSbar_2)| \Usc_2(S_1)\right\}| \Usc_1\right]\right)   - \\
        &\qquad \E\left(\mu_{11}(\bX)\E\left[\muhat_{12}(\bX, S_1)\Qschat_{12}(\bX, S_1) \} | \Usc_1\right]\right) +\\
        &\qquad \E\left(\mu_{11}(\bX)\E\left[\mu_{12}(\bX, S_1)\E\left\{\mu_{13}(\bX, \bSbar_2) - \muhat_{13}(\bX, \bSbar_2) | \Usc_2(S_1)\right\} |\Usc_1\right]  \right)\\
        &= \E\left(\mu_{11}(\bX)\Qsc_{11}(\bX) - \muhat_{11}(\bX)\Qschat_{11}(\bX) \right), 
\end{align*}
which, along with a similar result for $\varphi_0$ implies that 
\begin{align*}
    \E(\phihat) = \E\left\{\mu_{11}(\bX)\Qsc_{11}(\bX) - \mu_{01}(\bX)\Qsc_{01}(\bX)\right\} - \Deltat = 0.
\end{align*}

Next, we establish the conditions for the influence function of the residual treatment effect $\Deltastt$. Consider the influence function with estimated nuisance parameters. 
\begin{align*}
    \phihat_S &= \phi_S(\bO, \Deltastt, \Psihat) = 
\frac{G}{\ehat(\bX)}
                             \left(
                                \frac{A_1}{\gammahat_{11}(\bX)}\Qschat_{11}^*(\bX)\{Y_1 - \muhat_{11}(\bX)\} + \right.\\
                                &\qquad \frac{A_1Y_1A_2\pihat^*_1(\bX)}{\gammahat_{11}(\bX)\pihat_1(\bX, S_1)\gammahat_{12}(\bX, S_1)}\Qschat_{12}^*(\bX, S_1)\{Y_2 - \muhat_{12}(\bX, S_1)\} +\\
                                &\qquad \left. \frac{A_1Y_1A_2\pihat^*_1(\bX)Y_2A_3\pihat^*_2(\bX, S_1)}{\gammahat_{11}(\bX)\pihat_1(\bX, S_1)\gammahat_{12}(\bX, S_1)\pihat_2(\bX, \bSbar_2)\gammahat_{13}(\bX, \bSbar_2)}\{Y_3 - \muhat_{13}(\bX, \bSbar_2)\}\right) - \\
                                &\qquad \frac{1-G}{1-\ehat(\bX)}
                             \left(
                                \frac{A_1}{\gammahat_{01}(\bX)}\Qschat_{01}^*(\bX)\{Y_1 - \muhat_{01}(\bX)\} + \right.\\
                                &\qquad \frac{A_1Y_1A_2\{1-\pihat^*_1(\bX)\}}{\gammahat_{01}(\bX)\{1-\pihat_1(\bX, S_1)\}\gammahat_{02}(\bX, S_1)}\Qschat_{02}^*(\bX, S_1)\{Y_2 - \muhat_{02}(\bX, S_1)\} +\\
                                &\qquad \left. \frac{A_1Y_1A_2\{1-\pihat^*_1(\bX)\}Y_2A_3\{1-\pihat^*_2(\bX, S_1)\}}{\gammahat_{01}(\bX)\{1-\pihat_1(\bX, S_1)\}\gammahat_{02}(\bX, S_1)\{1-\pihat_2(\bX, \bSbar_2)\}\gammahat_{03}(\bX, \bSbar_2)}\{Y_3 - \muhat_{03}(\bX, \bSbar_2)\}\right) +\\
                                &\qquad\qquad \ehat(\bX)\inv \left(
                                    \frac{A_1Y_1}{\gammahat_{11}(\bX)}\pihat_1^*(\bX)\{\muhat_{12}(\bX, S_1)\Qschat^*_{12}(\bX, S_1) - \Qschat^*_{11}(\bX)\} \right. +\\
                                    &\left. \frac{A_1Y_1A_2Y_2\pihat_1^*(\bX)}{\gammahat_{11}(\bX)\pihat_1(\bX, S_1)\gammahat_{12}(\bX, S_1)}\pihat_2^*(\bX, S_1)\{\muhat_{13}(\bX, \bSbar_2) - \Qschat^*_{12}(\bX, S_1)\}
                                \right) - \\
                                &\qquad\qquad \{1-\ehat(\bX)\}\inv \left(
                                    \frac{A_1Y_1}{\gammahat_{01}(\bX)}\{1-\pihat_1^*(\bX)\}\{\muhat_{02}(\bX, S_1)\Qschat^*_{02}(\bX, S_1) - \Qschat^*_{01}(\bX)\} \right. +\\
                                    &\left. \frac{A_1Y_1A_2Y_2\{1-\pihat_1^*(\bX)\}}{\gammahat_{01}(\bX)\{1-\pihat_1(\bX, S_1)\}\gammahat_{02}(\bX, S_1)}\{1-\pihat_2^*(\bX, S_1)\}\{\muhat_{03}(\bX, \bSbar_2) - \Qschat^*_{02}(\bX, S_1)\}
                                \right) +\\
                                &\qquad\qquad \muhat_{11}\Qschat^*_{11} - \muhat_{01}\Qschat^*_{01} - \Deltastt \\
                                &= \varphi^*_{11} - \varphi^*_{01} + \varphi^*_{12} + \varphi^*_{02} + \muhat_{11}(\bX)\Qschat^*_{11}(\bX) - \muhat_{01}(\bX)\Qschat^*_{01}(\bX) - \Deltastt
    \end{align*}
    Now, we compute the expected value of the first term:
    \begin{align}\label{varphi_11}
        \E(\varphi^*_{11}) &= \E\left\{\frac{e(\bX)}{\ehat(\bX)}
                             \left(
                                \frac{\gamma_{11}(\bX)}{\gammahat_{11}(\bX)}\Qschat_{11}^*(\bX)\{\mu_{11}(\bX) - \muhat_{11}(\bX)\} + \right.\right.\\
                                &\qquad \frac{\gamma_{11}(\bX)\mu_{11}(\bX)\pihat^*_1(\bX)}{\gammahat_{11}(\bX)}\E\left[\frac{\gamma_{12}(\bX, S_1)}{\pihat_1(\bX, S_1)\gammahat_{12}(\bX, S_1)}\Qschat_{12}^*(\bX, S_1)\{\mu_{12}(\bX, S_1) - \muhat_{12}(\bX, S_1)\} | \Usc_1\right] +\\
                                &\qquad 
                                \frac{\gamma_{11}(\bX)\mu_{11}(\bX)\pihat^*_1(\bX)}{\gammahat_{11}(\bX)} \times\\
                                &\qquad 
                                    \E\left[\frac{\gamma_{12}(\bX, S_1)\mu_{12}(\bX, S_1)\pihat^*_2(\bX, S_1)}{\pihat_1(\bX, S_1)\gammahat_{12}(\bX, S_1)} \times \right.\\
                                    &\qquad \left. \left.\left.
                                        \E\left\{\frac{\gamma_{13}(\bX, \bSbar_2)}{\pihat_2(\bX, \bSbar_2)\gammahat_{13}(\bX, \bSbar_2)}\{\mu_{13}(\bX, \bSbar_2) - \muhat_{13}(\bX, \bSbar_2)\} | \Usc_2\right\} | \Usc_1\right]\right)\right\}
    \end{align}
    It is again straightforward to see that if $\muhat_{gj} = \mu_{gj}$, then $\E(\varphi^*_{11}) = 0$. Similarly, we can compute the expected value of the second treated-group term $\E(\varphi^*_{12})$.
    \begin{align}\label{varphi_12}
        \E(\varphi^*_{12}) &= \E\left\{\ehat(\bX)\inv \left(
                                    \frac{\gamma^*_{1}(\bX)\mu^*_{1}(\bX)}{\gammahat_{11}(\bX)}\pihat_1^*(\bX)\left[\E\left\{\muhat_{12}(\bX, S_1)\Qschat^*_{12}(\bX, S_1) | \Usc^*_1\right\}- \Qschat^*_{11}(\bX)\right] \right. \right.+\\
                                    &\left.\left. \frac{\gamma^*_{1}(\bX)\mu^*_{1}(\bX)\pihat_1^*(\bX)}{\gammahat_{11}(\bX)}
                                    \E\left[ \frac{\gamma^*_{12}(\bX, S_1)\mu^*_{12}(\bX, S_1)\pihat_2^*(\bX, S_1)}{\pihat_1(\bX, S_1)\gammahat_{12}(\bX, S_1)}\{\E\{\muhat_{13}(\bX, \bSbar_2)|\Usc_2^*\} - \Qschat^*_{12}(\bX, S_1)\} | \Usc^*_1\right]\right\}
                                \right)
    \end{align}
    where $\gamma^*_{k}(\bx, \bsbar_{k-1}) = \P(A_k = 1 | \bX = \bx, \bAbar_{k-1} = \bYbar_{k-1} = 1, \bSbar_{k-1} = \bsbar_{k-1})$ and $\mu^*_{k}(\bx, \bsbar_{k-1}) = \E(Y_k | \bX = \bx, \bAbar_{k} = \bYbar_{k-1} = 1, \bSbar_{k-1} = \bsbar_{k-1})$.
    Thus, if $\Qschat^*_{12}(\bX, S_1) = \E\{\muhat_{13}(\bX, \bSbar_2) | \Usc_2^*\}$ and $\Qschat^*_{11}(\bX) = \E\{\Qschat_{12}^*(\bX, S_1)\muhat_{12}(\bX, S_1)|\Usc_1^*\}$, then we obtain $\E(\varphi^*_{12}) = 0$.
    
    Now if $\gammahat_{gj} = \gamma_{gj}$, $\ehat = e$, $\pihat = \pi$ and $\pihat^* = \pi^*$, we have
    \begin{align*}
        \E(\varphi^*_{11}) &= \E\left\{
                                \Qschat_{11}^*(\bX)\{\mu_{11}(\bX) - \muhat_{11}(\bX)\} + \right.\\
                                &\qquad \mu_{11}(\bX)\E\left[\frac{\pi^*_1(\bX)}{\pi_1(\bX, S_1)}\Qschat_{12}^*(\bX, S_1)\{\mu_{12}(\bX, S_1) - \muhat_{12}(\bX, S_1)\} | \Usc_1\right] +\\
                                &\qquad \left.
                               \mu_{11}(\bX)
                                    \E\left[\frac{\mu_{12}(\bX, S_1)\pi^*_1(\bX)}{\pi_1(\bX, S_1)}
                                        \E\left\{\frac{\pi^*_2(\bX, S_1)}{\pi_2(\bX, \bSbar_2)}\{\mu_{13}(\bX, \bSbar_2) - \muhat_{13}(\bX, \bSbar_2)\} | \Usc_2\right\} | \Usc_1\right]\right\}
    \end{align*}
    Recall that $\pi_1, \pi_2, \pi^*_1, \pi^*_2$ are constructed so that
    \begin{align*}
        \E\left\{ \frac{\pi^*_2(\bX, S_1)}{\pi_2(\bX, \bSbar_2)} f(\bX, \bSbar_2) | \Usc_2\right\} &= \E\left\{ f(\bX, \bSbar_2) | \bX, A_1 = Y_1 = A_2, Y_2 = 1, S_1\right\} = \E\left\{ f(\bX, \bSbar_2) | \Usc_2^*\right\}\\
        \E\left\{ \frac{\pi^*_1(\bX)}{\pi_1(\bX, S_1)} f(\bX, S_1) | \Usc_1\right\} &= \E\left\{ f(\bX, \bSbar_1) | \bX, A_1 = Y_1 = 1\right\} = \E\left\{ f(\bX, \bSbar_1) | \Usc_1^*\right\},
    \end{align*}
    so
    \begin{align*}
        \E(\varphi^*_{11}) &= \E\left\{
                                \Qschat_{11}^*(\bX)\{\mu_{11}(\bX) - \muhat_{11}(\bX)\} + \right.\\
                                &\qquad \mu_{11}(\bX)\E\left[\frac{\pi^*_1(\bX)}{\pi_1(\bX, S_1)}\Qschat_{12}^*(\bX, S_1)\{\mu_{12}(\bX, S_1) - \muhat_{12}(\bX, S_1)\} | \Usc_1\right] +\\
                                &\qquad \left.
                               \mu_{11}(\bX)
                                    \E\left[\frac{\mu_{12}(\bX, S_1)\pi^*_1(\bX)}{\pi_1(\bX, S_1)}
                                        \E\left\{\frac{\pi^*_2(\bX, S_1)}{\pi_2(\bX, \bSbar_2)}\{\mu_{13}(\bX, \bSbar_2) - \muhat_{13}(\bX, \bSbar_2)\} | \Usc_2\right\} | \Usc_1\right]\right\}\\
                                        &= \E\left\{
                                \Qschat_{11}^*(\bX)\{\mu_{11}(\bX) - \muhat_{11}(\bX)\} + \right.\\
                                &\qquad \mu_{11}(\bX)\E\left[\Qschat_{12}^*(\bX, S_1)\{\mu_{12}(\bX, S_1) - \muhat_{12}(\bX, S_1)\} | \Usc^*_1\right] +\\
                                &\qquad \left.
                               \mu_{11}(\bX)
                                    \E\left[\mu_{12}(\bX, S_1)
                                        \E\left\{\{\mu_{13}(\bX, \bSbar_2) - \muhat_{13}(\bX, \bSbar_2)\} | \Usc^*_2\right\} | \Usc^*_1\right]\right\}
    \end{align*}
    We can perform similar calculations for $\varphi^*_{12}$ when $\gammahat_{gj} = \gamma_{gj}$, $\ehat = e$, $\pihat = \pi$ and $\pihat^* = \pi^*$
    \begin{align*}
        \E(\varphi^*_{12}) &= \E\left\{\ehat(\bX)\inv \left(
                                    \frac{\gamma^*_{1}(\bX)\mu^*_{1}(\bX)}{\gammahat_{11}(\bX)}\pihat_1^*(\bX)\left[\E\left\{\muhat_{12}(\bX, S_1)\Qschat^*_{12}(\bX, S_1) | \Usc^*_1\right\}- \Qschat^*_{11}(\bX)\right] \right. \right.+\\
                                    &\left.\left. \frac{\gamma^*_{1}(\bX)\mu^*_{1}(\bX)\pihat_1^*(\bX)}{\gammahat_{11}(\bX)}
                                    \E\left[ \frac{\gamma^*_{12}(\bX, S_1)\mu^*_{12}(\bX, S_1)\pihat_2^*(\bX, S_1)}{\pihat_1(\bX, S_1)\gammahat_{12}(\bX, S_1)}\{\E\{\muhat_{13}(\bX, \bSbar_2)|\Usc_2^*\} - \Qschat^*_{12}(\bX, S_1)\} | \Usc^*_1\right]\right\}
                                \right)
    \end{align*}
    where $\gamma^*_{k}(\bx, \bsbar_{k-1}) = \P(A_k = 1 | \bX = \bx, \bAbar_{k-1} = \bYbar_{k-1} = 1, \bSbar_{k-1} = \bsbar_{k-1})$ and $\mu^*_{k}(\bx, \bsbar_{k-1}) = \E(Y_k | \bX = \bx, \bAbar_{k} = \bYbar_{k-1} = 1, \bSbar_{k-1} = \bsbar_{k-1})$.
    Thus, if $\Qschat^*_{12}(\bX, S_1) = \E\{\muhat_{13}(\bX, \bSbar_2) | \Usc_2^*\}$ and $\Qschat^*_{11}(\bX) = \E\{\Qschat_{12}^*(\bX, S_1)\muhat_{12}(\bX, S_1)|\Usc_1^*\}$, then we obtain $\E(\varphi^*_{12}) = 0$. Alternatively, if $\ehat = e, \pihat^*_k = \pi^*_k, \pihat_k = \pi_k$, and $\gammahat_{1k} = \gamma_{1k}$, we have 
    \begin{align*}
        \E(\varphi^*_{12}) &= \E\left\{e(\bX)\inv \left(
                                    \frac{\gamma^*_{1}(\bX)\mu^*_{1}(\bX)}{\gamma_{11}(\bX)}\pi_1^*(\bX)\left[\E\left\{\muhat_{12}(\bX, S_1)\Qschat^*_{12}(\bX, S_1) | \Usc^*_1\right\}- \Qschat^*_{11}(\bX)\right] \right. \right.+\\
                                    &\left.\left. \frac{\gamma^*_{1}(\bX)\mu^*_{1}(\bX)\pi_1^*(\bX)}{\gamma_{11}(\bX)}
                                    \E\left[ \frac{\gamma^*_{2}(\bX, S_1)\mu^*_{2}(\bX, S_1)\pi_2^*(\bX, S_1)}{\pi_1(\bX, S_1)\gamma_{12}(\bX, S_1)}\{\E\{\muhat_{13}(\bX, \bSbar_2)|\Usc_2^*\} - \Qschat^*_{12}(\bX, S_1)\} | \Usc^*_1\right]\right\}
                                \right).
    \end{align*}
    Now note that 
    \begin{align*}
        &\frac{\gamma^*_{2}(\bX, S_1)\mu^*_{2}(\bX, S_1)\pi_2^*(\bX, S_1)}{\pi_1(\bX, S_1)\gamma_{12}(\bX, S_1)} = \\
        &\qquad \frac{\P(A_2 = 1 | \bX, A_1 = Y_1 = 1, S_1)\E(Y_2 | \bX, 
        \bAbar_2 = Y_1 = 1, S_1)\P(G = 1 | \bX, \bAbar_2 = \bYbar_2 = 1, S_1)}{\P(G = 1 | \bX, A_1 = Y_1, S_1 = 1)\P(A_2 = 1 | \bX, G = 1, A_1 = Y_1 = 1, S_1)} \\
        &\qquad \frac{\P(A_2 = 1 | \bX, A_1 = Y_1 = 1, S_1)\E(Y_2 | \bX, 
        \bAbar_2 = Y_1 = 1, S_1)\P(A_2 = Y_2 = 1 | \bX, G = 1, A_1 = Y_1 = 1, S_1)}{\P(A_2 = 1 | \bX, G = 1, A_1 = Y_1 = 1, S_1)\P(A_2 = Y_2 = 1 | \bX, A_1 = Y_1 = 1, S_1)} \\
        &= \mu_{12}(\bX, S_1)
    \end{align*}
and similarly
 \begin{align*}
     \frac{\gamma^*_{1}(\bX)\mu^*_{1}(\bX)\pi_1^*(\bX)}{e(\bX)\gamma_{11}(\bX)} = \mu_{11}(\bX),
 \end{align*}
 so 
 \begin{align*}
        \E(\varphi^*_{12}) &= \E\left\{
                                    \mu_{11}(\bX)\left[\E\left\{\muhat_{12}(\bX, S_1)\Qschat^*_{12}(\bX, S_1) | \Usc^*_1\right\}- \Qschat^*_{11}(\bX)\right] \right. +\\
                                    &\left.\left. \mu_{11}(\bX)
                                    \E\left[ \mu_{12}(\bX, S_1)\{\E\{\muhat_{13}(\bX, \bSbar_2)|\Usc_2^*\} - \Qschat^*_{12}(\bX, S_1)\} | \Usc^*_1\right]\right\}
                                \right).
    \end{align*}
    Putting these two terms together, we have that
    \begin{align*}
        \E(\varphi^*_{11} + \varphi^*_{12}) = -\E\left\{\muhat_{11}(\bX)\Qschat^*_{11}(\bX)\right\}.
    \end{align*}
    Along with the corresponding result for $\E(\varphi^*_{01} + \varphi^*_{02})$, this implies $\E(\phihat_S) = 0.$

\subsection{Asymptotic distribution}
The argument for the asymptotic normality of the one-step estimator follows standard arguments for similar cross-fitting estimators \citep{kennedy2022semiparametric}. First, recall that due to sample-splitting $\Deltahat_\bS(t, t_0) = \frac{1}{2}\Deltahat_{\bS0} + \frac{1}{2}\Deltahat_{\bS1}, \Deltahat(t) = \frac{1}{2}\Deltahat_0 + \frac{1}{2}\Deltahat_1$ for 
\begin{align*}
    \Deltahat_{\bS\ell} = \frac{2}{n}\sum_{i \in \Isc_\ell}\phitilde_\bS(\bO_i, \Psihat_{1-\ell})\\
    \Deltahat_{\ell} = \frac{2}{n}\sum_{i \in \Isc_\ell}\phitilde(\bO_i, \Psihat_{1-\ell}).
\end{align*}
Considering only $\Deltastt$ (since the argument for $\Deltat$ is identical), we have
\begin{align*}
\Deltahat_\bS(t, t_0) - \Deltastt &= \sum_{\ell=0}^1\E^\ell_n\{\phitilde_\bS(\bO, \Psihat_{1-\ell})\} - \E\{\phitilde_\bS(\bO, \Psi)\}\\
&= (\E_n - \E)\{\phitilde_\bS(\bO, \Psi)\} + \sum_{\ell=0}^1(\E^\ell_n - \E)\{\phitilde_\bS(\bO, \Psihat_{1-\ell}) - \phitilde_\bS(\bO, \Psi)\} + \\
&\qquad \sum_{\ell=0}^1\E\{\phitilde_\bS(\bO, \Psihat_{1-\ell}) - \phitilde_\bS(\bO, \Psi)\}\\
&= (\E_n - \E)\{\phitilde_\bS(\bO, \Psi)\} + \sum_{\ell = 1}^{2} (T_{1\ell} + T_{2\ell})
\end{align*}
for $\E_n\{f(\bO)\} = n\inv\sumin f(\bO_i)$ and $\E^\ell_n\{f(\bO)\} = \frac{2}{n}\sum_{i \in \Isc_\ell} f(\bO_i)$ for any $f$.
We will show that the remaining terms converge to 0 fast enough to be ignored asymptotically. First, note that $|T_{1\ell}| = o_p(\nnhalf)$ because of Assumptions \ref{asmp:pos2} and \ref{asmp:censpos2} and equation \eqref{convergence-rates} which yield the result based on Proposition 1 of \citet{kennedy2022semiparametric}. 

Next, consider the treated-group portion of the term $T_{2\ell} = \E\{\phitilde_\bS(\bO, \Psihat_{1-\ell}) - \phitilde_\bS(\bO, \Psi)\}$, which we can obtain from \eqref{varphi_11} and \eqref{varphi_12}, as $\iota_1 + \iota_2 + \iota_3 + \iota_4 + \iota_5 + \E\left\{\muhat_{11}(\bX)\Qschat_{11}^*(\bX) - \mu_{11}(\bX)\Qsc^*_{11}(\bX)\right\}$ for
\begin{align*}
    \iota_1 = \E\left[\frac{e(\bX)}{\ehat(\bX)}\frac{\gamma_{11}(\bX)}{\gammahat_{11}(\bX)} \Qschat^*_{11}(\bX)\{\mu_{11}(\bX) - \muhat_{11}(\bX)\} \right],
    \end{align*}
\begin{align*}
    \iota_2 &= \E\left[\frac{e(\bX)}{\ehat(\bX)}\frac{\gamma_{11}(\bX)}{\gammahat_{11}(\bX)}\mu_{11}(\bX)\E\left[\left\{\frac{\pihat^*_1(\bX)}{\pihat_1(\bX, S_1)} - \frac{\pi^*_1(\bX)}{\pi_1(\bX, S_1)}\right\}\frac{\gamma_{12}(\bX, S_1)}{\gammahat_{12}(\bX, S_1)}\Qschat_{12}^*(\bX, S_1)\{\mu_{12}(\bX, S_1) - \muhat_{12}(\bX, S_1)\} | \Usc_1\right] \right] + \\
    &\qquad \E\left[\frac{e(\bX)}{\ehat(\bX)}\frac{\gamma_{11}(\bX)}{\gammahat_{11}(\bX)}\mu_{11}(\bX)\E\left[\frac{\gamma_{12}(\bX, S_1)}{\gammahat_{12}(\bX, S_1)}\Qschat_{12}^*(\bX, S_1)\{\mu_{12}(\bX, S_1) - \muhat_{12}(\bX, S_1)\} | \Usc^*_1\right] \right],
\end{align*}
\begin{align*}
    \iota_3 &= \E\left(\frac{e(\bX)}{\ehat(\bX)}\frac{\gamma_{11}(\bX)}{\gammahat_{11}(\bX)}\mu_{11}(\bX)
    \E\left[\frac{\pihat^*_1(\bX)}{\pihat_1(\bX, S_1)}\frac{\gamma_{12}(\bX, S_1)}{\gammahat_{12}(\bX, S_1)}\mu_{12}(\bX, S_1) \times \right.\right.\\
                                    &\qquad \left. \left.
                                        \E\left\{\frac{\pihat^*_2(\bX, S_1)}{\pihat_2(\bX, \bSbar_2)}\frac{\gamma_{13}(\bX, \bSbar_2)}{\gammahat_{13}(\bX, \bSbar_2)}\{\mu_{13}(\bX, \bSbar_2) - \muhat_{13}(\bX, \bSbar_2)\} | \Usc_2\right\} | \Usc_1\right]\right)\\
    &= \E\left(\frac{e(\bX)}{\ehat(\bX)}\frac{\gamma_{11}(\bX)}{\gammahat_{11}(\bX)}\mu_{11}(\bX)
    \E\left[\left\{\frac{\pihat^*_1(\bX)}{\pihat_1(\bX, S_1)} - \frac{\pi^*_1(\bX)}{\pi_1(\bX, S_1)}\right\}\frac{\gamma_{12}(\bX, S_1)}{\gammahat_{12}(\bX, S_1)}\mu_{12}(\bX, S_1) \times \right.\right.\\
                                    &\qquad \left. \left.
                                        \E\left\{\frac{\pihat^*_2(\bX, S_1)}{\pihat_2(\bX, \bSbar_2)}\frac{\gamma_{13}(\bX, \bSbar_2)}{\gammahat_{13}(\bX, \bSbar_2)}\{\mu_{13}(\bX, \bSbar_2) - \muhat_{13}(\bX, \bSbar_2)\} | \Usc_2\right\} | \Usc_1\right]\right) + \\
    &\qquad \E\left(\frac{e(\bX)}{\ehat(\bX)}\frac{\gamma_{11}(\bX)}{\gammahat_{11}(\bX)}\mu_{11}(\bX)
    \E\left[\frac{\gamma_{12}(\bX, S_1)}{\gammahat_{12}(\bX, S_1)}\mu_{12}(\bX, S_1) \times \right. \right.\\
                                    &\qquad \left. \left.
                                        \E\left\{\left[\frac{\pihat^*_2(\bX, S_1)}{\pihat_2(\bX, \bSbar_2)} - \frac{\pi^*_2(\bX, S_1)}{\pi_2(\bX, \bSbar_2)} \right]\frac{\gamma_{13}(\bX, \bSbar_2)}{\gammahat_{13}(\bX, \bSbar_2)}\{\mu_{13}(\bX, \bSbar_2) - \muhat_{13}(\bX, \bSbar_2)\} | \Usc_2\right\} | \Usc^*_1\right]\right) + \\
    &\qquad \E\left(\frac{e(\bX)}{\ehat(\bX)}\frac{\gamma_{11}(\bX)}{\gammahat_{11}(\bX)}\mu_{11}(\bX)
    \E\left[\frac{\gamma_{12}(\bX, S_1)}{\gammahat_{12}(\bX, S_1)}\mu_{12}(\bX, S_1)
                                        \E\left\{\frac{\gamma_{13}(\bX, \bSbar_2)}{\gammahat_{13}(\bX, \bSbar_2)}\{\mu_{13}(\bX, \bSbar_2) - \muhat_{13}(\bX, \bSbar_2)\} | \Usc^*_2\right\} | \Usc^*_1\right]\right),
\end{align*}
\begin{align*}
    \iota_4 &= \E\left\{
                                    \frac{\gamma^*_{1}(\bX)\mu^*_{1}(\bX)\pihat_1^*(\bX)}{\ehat(\bX)\gammahat_{11}(\bX)}\left[\E\left\{\muhat_{12}(\bX, S_1)\Qschat^*_{12}(\bX, S_1) | \Usc^*_1\right\}- \Qschat^*_{11}(\bX)\right]  \right\}\\
    &= \E\left\{\left[\frac{\gamma^*_{1}(\bX)\mu^*_{1}(\bX)\pihat_1^*(\bX)}{\ehat(\bX)\gammahat_{11}(\bX)} -\frac{\gamma^*_{1}(\bX)\mu^*_{1}(\bX)\pi_1^*(\bX)}{e(\bX)\gamma_{11}(\bX)} \right]
                                    \left[\E\left\{\muhat_{12}(\bX, S_1)\Qschat^*_{12}(\bX, S_1) | \Usc^*_1\right\}- \Qschat^*_{11}(\bX)\right]  \right. + \\
                                    &\qquad \left.\mu_{11}(\bX)\left[\E\left\{\muhat_{12}(\bX, S_1)\Qschat^*_{12}(\bX, S_1) | \Usc^*_1\right\}- \Qschat^*_{11}(\bX)\right] \right\},
\end{align*}
\begin{align*}
    \iota_5 &= \E\left\{\left[\frac{\gamma^*_{1}(\bX)\mu^*_{1}(\bX)\pihat_1^*(\bX)}{\ehat(\bX)\gammahat_{11}(\bX)} - \frac{\gamma^*_{1}(\bX)\mu^*_{1}(\bX)\pi_1^*(\bX)}{e(\bX)\gamma_{11}(\bX)}\right]
                                    \E\left[ \frac{\gamma^*_{12}(\bX, S_1)\mu^*_{12}(\bX, S_1)\pihat_2^*(\bX, S_1)}{\pihat_1(\bX, S_1)\gammahat_{12}(\bX, S_1)}\{\E\{\muhat_{13}(\bX, \bSbar_2)|\Usc_2^*\} - \Qschat^*_{12}(\bX, S_1)\} | \Usc^*_1\right]\right\} + \\
&\qquad \E\left\{\mu_{11}(\bX)
                                    \E\left[ \left\{\frac{\gamma^*_{12}(\bX, S_1)\mu^*_{12}(\bX, S_1)\pihat_2^*(\bX, S_1)}{\pihat_1(\bX, S_1)\gammahat_{12}(\bX, S_1)} - \frac{\gamma^*_{12}(\bX, S_1)\mu^*_{12}(\bX, S_1)\pi_2^*(\bX, S_1)}{\pi_1(\bX, S_1)\gamma_{12}(\bX, S_1)}\right\}\{\E\{\muhat_{13}(\bX, \bSbar_2)|\Usc_2^*\} - \Qschat^*_{12}(\bX, S_1)\} | \Usc^*_1\right]\right\} + \\
&\qquad \E\left\{\mu_{11}(\bX)\E\left[ \mu_{12}(\bX, S_1)\{\E\{\muhat_{13}(\bX, \bSbar_2)|\Usc_2^*\} - \Qschat^*_{12}(\bX, S_1)\} | \Usc^*_1\right]\right\}.
\end{align*}

Therefore, we can write the treated-group part of $T_{2\ell}$ as
\begin{align*}
    &\iota_1 + \iota_2 +\iota_3 + \iota_4 + \iota_5+ \E\left\{\muhat_{11}(\bX)\Qschat_{11}^*(\bX) - \mu_{11}(\bX)\Qsc^*_{11}(\bX)\right\} =\\
    &\E\left[\frac{\left\{e(\bX)\gamma_{11}(\bX) - \ehat(\bX)\gammahat_{11}(\bX) \right\}}{\ehat(\bX)\gammahat_{11}(\bX)}\Qschat^*_{11}(\bX)\left\{\mu_{11}(\bX) - \muhat_{11}(\bX)\right\}\right] + \\
    &\E\left(\E\left[\frac{\left\{e(\bX)\gamma_{11}(\bX)\gamma_{12}(\bX, S_1) - \ehat(\bX)\gammahat_{11}(\bX)\gammahat_{12}(\bX, S_1) \right\}}{\ehat(\bX)\gammahat_{11}(\bX)\gammahat_{12}(\bX, S_1)}\mu_{11}(\bX)\Qschat^*_{12}(\bX)\left\{\mu_{12}(\bX, S_1) - \muhat_{12}(\bX, S_1)\right\}| \Usc_1^*\right]\right) +\\
    &\E\left\{\left(\E\left[\frac{\left\{e(\bX)\gamma_{11}(\bX)\gamma_{12}(\bX, S_1)\gamma_{13}(\bX, \bSbar_2) - \ehat(\bX)\gammahat_{11}(\bX)\gammahat_{12}(\bX, S_1)\gammahat_{13}(\bX, \bSbar_2) \right\}}{\ehat(\bX)\gammahat_{11}(\bX)\gammahat_{12}(\bX, S_1)\gammahat_{13}(\bX, \bSbar_2)}\mu_{11}(\bX)\mu_{12}(\bX, S_1)\left\{\mu_{13}(\bX, \bSbar_2) - \muhat_{13}(\bX, \bSbar_2)\right\}| \Usc_2^*\right]| \Usc_1^*\right)\right\} +\\
    &\E\left[\frac{e(\bX)}{\ehat(\bX)}\frac{\gamma_{11}(\bX)}{\gammahat_{11}(\bX)}\mu_{11}(\bX)\E\left[\left\{\frac{\pihat^*_1(\bX)}{\pihat_1(\bX, S_1)} - \frac{\pi^*_1(\bX)}{\pi_1(\bX, S_1)}\right\}\frac{\gamma_{12}(\bX, S_1)}{\gammahat_{12}(\bX, S_1)}\Qschat_{12}^*(\bX, S_1)\{\mu_{12}(\bX, S_1) - \muhat_{12}(\bX, S_1)\} | \Usc_1\right] \right] + \\
    &\E\left(\frac{e(\bX)}{\ehat(\bX)}\frac{\gamma_{11}(\bX)}{\gammahat_{11}(\bX)}\mu_{11}(\bX)
    \E\left[\left\{\frac{\pihat^*_1(\bX)}{\pihat_1(\bX, S_1)} - \frac{\pi^*_1(\bX)}{\pi_1(\bX, S_1)}\right\}\frac{\gamma_{12}(\bX, S_1)}{\gammahat_{12}(\bX, S_1)}\mu_{12}(\bX, S_1) \times \right.\right.\\
                                    &\qquad \left.\left.
                                        \E\left\{\frac{\pihat^*_2(\bX, S_1)}{\pihat_2(\bX, \bSbar_2)}\frac{\gamma_{13}(\bX, \bSbar_2)}{\gammahat_{13}(\bX, \bSbar_2)}\{\mu_{13}(\bX, \bSbar_2) - \muhat_{13}(\bX, \bSbar_2)\} | \Usc_2\right\} | \Usc_1\right]\right) + \\
    &\qquad \E\left(\frac{e(\bX)}{\ehat(\bX)}\frac{\gamma_{11}(\bX)}{\gammahat_{11}(\bX)}\mu_{11}(\bX)
    \E\left[\frac{\gamma_{12}(\bX, S_1)}{\gammahat_{12}(\bX, S_1)}\mu_{12}(\bX, S_1) \times \right.\right.\\
                                    &\qquad \left. \left.
                                        \E\left\{\left[\frac{\pihat^*_2(\bX, S_1)}{\pihat_2(\bX, \bSbar_2)} - \frac{\pi^*_2(\bX, S_1)}{\pi_2(\bX, \bSbar_2)} \right]\frac{\gamma_{13}(\bX, \bSbar_2)}{\gammahat_{13}(\bX, \bSbar_2)}\{\mu_{13}(\bX, \bSbar_2) - \muhat_{13}(\bX, \bSbar_2)\} | \Usc_2\right\} | \Usc^*_1\right]\right) + \\
    &\E\left\{\left[\frac{\gamma^*_{1}(\bX)\mu^*_{1}(\bX)\pihat_1^*(\bX)}{\ehat(\bX)\gammahat_{11}(\bX)} -\frac{\gamma^*_{1}(\bX)\mu^*_{1}(\bX)\pi_1^*(\bX)}{e(\bX)\gamma_{11}(\bX)} \right]
                                    \left[\E\left\{\muhat_{12}(\bX, S_1)\Qschat^*_{12}(\bX, S_1) | \Usc^*_1\right\}- \Qschat^*_{11}(\bX)\right] \right\} + \\
    &\E\left\{\left[\frac{\gamma^*_{1}(\bX)\mu^*_{1}(\bX)\pihat_1^*(\bX)}{\ehat(\bX)\gammahat_{11}(\bX)} - \frac{\gamma^*_{1}(\bX)\mu^*_{1}(\bX)\pi_1^*(\bX)}{e(\bX)\gamma_{11}(\bX)}\right]
                                    \E\left[ \frac{\gamma^*_{12}(\bX, S_1)\mu^*_{12}(\bX, S_1)\pihat_2^*(\bX, S_1)}{\pihat_1(\bX, S_1)\gammahat_{12}(\bX, S_1)}\{\E\{\muhat_{13}(\bX, \bSbar_2)|\Usc_2^*\} - \Qschat^*_{12}(\bX, S_1)\} | \Usc^*_1\right]\right\} + \\
&\qquad \E\left\{\mu_{11}(\bX)
                                    \E\left[ \left\{\frac{\gamma^*_{12}(\bX, S_1)\mu^*_{12}(\bX, S_1)\pihat_2^*(\bX, S_1)}{\pihat_1(\bX, S_1)\gammahat_{12}(\bX, S_1)} - \frac{\gamma^*_{12}(\bX, S_1)\mu^*_{12}(\bX, S_1)\pi_2^*(\bX, S_1)}{\pi_1(\bX, S_1)\gamma_{12}(\bX, S_1)}\right\}\{\E\{\muhat_{13}(\bX, \bSbar_2)|\Usc_2^*\} - \Qschat^*_{12}(\bX, S_1)\} | \Usc^*_1\right]\right\}
\end{align*}
Along with a similar calculation for the control group portion of $T_{2\ell}$, this yields the result in the main text because
\begin{align*}
\Deltahat_\bS(t, t_0) - \Deltastt &= (\E_n - \E)\{\phitilde_\bS(\bO, \Psi)\} + o_p(\nnhalf),\\
\Deltahat(t) - \Deltat &= (\E_n - \E)\{\phitilde(\bO, \Psi)\} + o_p(\nnhalf).
\end{align*}

\section{Simulation Details \label{app:sims}}
\subsection{Comparison to joint model approach}

A general approach to estimate the PTE using a joint model, as proposed in \citet{taylor2002surrogate}, can be described as follows. Let $G$ denote the treatment indicator, $S^*$ denote the true longitudinal marker, and $S$ denote the observed surrogate marker. Their specified models are
$$S_{i}(t_j) = S^*_{i}(t_j) + e_{ij} = intercept + slope*t_j + \beta G_i + e_{ij}$$
$$\lambda_i(t) = \lambda_0(t)\exp \{\gamma S^*_i(t)+ \omega G_i \}.$$
Then the PTE can be defined as: $PTE = \frac{\beta \gamma}{\beta \gamma + \omega}$. 
 We estimate the this PTE using the \texttt{JM} package as follows. Let \texttt{data.long} denote the data in long form with information on the longitudinal marker; \texttt{y} in \texttt{data.long} is $S$, the measured marker at time $t_j$ denoted as \texttt{month}, \texttt{g} is the treatment indicator and \texttt{id} is the unique id for an individual. Let \texttt{data.surv} denote the data with 1 row per individual with information on the event time; \texttt{fup} is the observed event/censoring time and \texttt{event} is the event indicator. The R code is:\\
 
\noindent \texttt{>lmeFit <- lme(y $\sim$ month + g, random = ~ 1 | id, data = data.long)}\\
\texttt{>coxFit <- coxph(Surv(fup, event) $\sim$ g, data = data.surv, x = TRUE)}\\
\texttt{>jointFit <- jointModel(lmeFit, coxFit, timeVar = "month", method = "weibull-AFT-GH")}\\
\texttt{>summary(jointFit)}\\
\texttt{>summary(coxFit)}\\
\texttt{>summary(lmeFit)}\\

\noindent \texttt{\#pte as defined by Taylor and Wang 2002}\\
\texttt{>pte = (jointFit\$coefficient\$betas[3]*jointFit\$coefficient\$alpha)}\\
\texttt{+/(jointFit\$coefficient\$betas[3]*jointFit\$coefficient\$alpha+}\\
\texttt{+jointFit\$coefficient\$gammas[2])}

\subsection{Simulation setup}

In all settings, we let $X_i \sim N(0, 1)$ be a pre-treatment covariate, and $G_i$ be the treatment indicator generated as a Bernoulli with probability given by $e(X_i) = \text{expit}(X_i) = e^{X_i}/(1+e^{X_i})$. Next, for $k = 1, ...,5$, the counterfactual surrogates were $S_{ik}\supg \sim N(\{\zeta_{gk}(X_i, S\supg_{ik-1})\},1)$, where $\zeta_{gk}(X_i, S\supg_{ik-1}) = \alpha_0g+\alpha_1X_i + \alpha_2S\supg_{ik-1}$. For $k = 1, ..., 6$, the counterfactual outcomes were $Y_{ik}\supg \sim \text{Bernoulli}\{\mu_{gk}(X_i, S\supg_{ik-1})\}$, with $\mu_{gk}(X_i, S\supg_{ik-1}) = \text{expit}\{\alpha_3 + \alpha_4g + \alpha_5S_{ik-1}\supg + \alpha_6gS_{ik-1}\supg + \alpha_7X_i\}.$ Censoring was generated from an Exponential(0.1) distribution. In Setting 1, $\balpha=(\alpha_0,\alpha_1,\alpha_2,\alpha_3, \alpha_4, \alpha_5, \alpha_6, \alpha_7)' = (-0.1, 0.5, 0.25, -2, -1, 0.5, 0, 0.3)$ such that $R_S(t, t_0) = 0.028$, and 29\% of observations are censored before $t$. In Setting 2, $\mathbf{\alpha} = (-0.5, 0.5, 0.25, -5, -0.05, 4.5, -0.05, 0.3)$ such that  $R_S(t, t_0) = 0.966$ and 33\% of observations are censored before $t$. In Setting 3, $\mathbf{\alpha}=(-0.5, 0.5, 0.25, -5, -1, 4, -0.1, 0.3)$ such that $R_S(t, t_0) = 0.608$ and 33\% of observations are censored before $t$. For both estimators, the estimated variances were obtained as the variances of the empirical influence functions from the plug-in estimator and confidence intervals were constructed using a normal approximation.

\section{Selecting the optimal $t_0$ \label{app:test}}

As mentioned in the main text, one approach to selecting $t_0$ is to find the earliest $t^*$ such that the PTE has not declined too much from $t_\Lsc$. Stating it more formally, we could select $t_0$ to be the minimum $t^* \in \Omega = \{1,2, ..., t_\Lsc\}$ such that, for a given $\epsilon > 0$,
\begin{align*}
    R_\bS(t, t^*) > R_\bS(t, t_\Lsc) - \epsilon.
\end{align*}

We can find this minimizing $t_0$ by repeatedly testing the one-sided null hypothesis $H_{0j}: R_\bS(t, j) = R_\bS(t, t_\Lsc) - \epsilon$ against the alternative $H_{1j}: R_\bS(t, j) < R_\bS(t, t_\Lsc) - \epsilon$ for $j \in \Omega$. Rejecting this null hypothesis would amount to concluding that the proportion of treatment explained by the surrogate up to time $j$ is more than $\epsilon$ smaller than the PTE by the surrogate at $t_\Lsc$. We thus would not feel comfortable using the surrogate measured at time $j$ in place of the surrogate measured at time $t_\Lsc$.

Testing $H_{0j}: R_\bS(t, j) = R_\bS(t, t_\Lsc) - \epsilon$ individually can be based on the quantity
\begin{align*}
    \deltahat_j = \Rhat_\bS(t, t_\Lsc) - \Rhat_\bS(t, j) - \epsilon = \frac{\Deltahat_\bS(t,j) - \Deltahat_\bS(t,t_\Lsc)}{\Deltahat(t)} - \epsilon.
\end{align*}
Note that the form of $\deltahat_j$ is very similar to the form of $\Rhat_S(t, t_0)$, and it can be shown that:
\begin{align*}
    \sqrt{n}\{\deltahat_j - \delta_j\} \rightarrow N(0, \sigma_\delta^{2}), \quad \mbox{where }
    \sigma_\delta^2 = \bzeta\trans\Sigma\bzeta,
\end{align*} 
$\bzeta = (\Deltat\inv, -\Deltat\inv, -\delta_j\Deltat^{-2})$, $\Sigma = \E(\bphi\bphi\trans)$, and 
\begin{align*}
\bphi = \left[\phi_S\{\bO, \Delta_S(t, j), \bPsi\}, \phi_S\{\bO, \Delta_S(t, t_\Lsc), \bPsi\}, \phi(\bO, \Deltat, \bPsi\}\right],
\end{align*}
and $\delta_j =  R_\bS(t, t_\Lsc) - R_\bS(t, j) - \epsilon.$
Therefore, under $H_{0j}$, $\deltahat_j$ is approximately distributed as $N(0, \sigma^2_\delta/n)$. For a level $\alpha$ test, we may reject $H_{0j}$ if 
\begin{align}\label{solo-test}
    \tau_j = \sqrt{n}\deltahat_j/\sigma_\delta > z_{1-\alpha}
\end{align}
where $z_{1-\alpha}$ is the $(1-\alpha)$ quantile of the standard normal distribution.

One can construct a stepdown testing procedure \citep{romano2005exact} to ensure familywise error rate (FWER) control over this set of tests, meaning that we ensure that the probability of falsely rejecting any of these $t_\Lsc-1$ tests is maintained at a given error level $\alpha$. That is, one can first test the intersection hypothesis that all null hypotheses ($H_{01}, ..., H_{0\Lsc-1}$) are true
using the maximum test statistic among all hypotheses:
\begin{align*}
    T^{(1)} = \max_{j \in \Omega} \tau_j
\end{align*}
and reject this intersection null hypothesis if $T^{(1)} > c_{n, \Lsc-1}(\alpha)$, where $c_{n, \Omega}(\alpha)$ is the level $\alpha$ critical value for testing the intersection null hypothesis. This critical value must be obtained using a resampling procedure that takes into account the joint distribution of the test statistics and accounts for the fact that the test statistics are likely highly correlated to each other. However, in many cases, a monotonicity assumption may be warranted and can greatly simplify the testing procedure. Under the assumption that adding additional surrogate information never decreases the PTE -- i.e., $R_S(t, t_j) \leq R_S(t, t_k)$ for $t_k > t_j$, one can conclude that $T^{(k)} = t_k$ for $k = 1, ..., \Lsc-1$, and the critical value $c_{n, \Omega_k}$ may be selected as $z_{1-\alpha}$ as in \eqref{solo-test}. 

If one fails to reject, then the procedure is stopped and the conclusion would be that every timepoint is no more than $\epsilon$ worse than the full $t_\Lsc$ timepoints in terms of PTE. Thus, it be reasonable to consider  collecting the surrogate at only the first timepoint. If we do reject, the timepoint corresponding to $T^{(1)}$ -- the $j^*$ where $\tau_{j^*} = T^{(1)}$ -- is removed from the set of null hypotheses, and a new set $\Omega_2 = \Omega \setminus j^*$ is constructed, and the process is repeated.

The full procedure proceeds as follows. Let $\Omega_1 = \Omega$. For $k = 1, ..., \Lsc-1$
\begin{enumerate}
    \item Construct the maximum test statistic over $\Omega_k: T^{(k)} = \max_{j \in \Omega_k} \tau_j$.
    \item Compute the critical value $c_{n, \Omega_k}(\alpha)$ using resampling.
    \item If $T^{(k)} < c_{n, \Omega_k}(\alpha)$, stop and recommend collecting the surrogate through $t^* = \min \Omega_k$. 
    \item If $T^{(k)} > c_{n, \Omega_k}(\alpha)$, set $\Omega_{k+1} = \Omega_k \setminus j^*$, where $\tau_{j^*} = T^{(k)}$. Return to step 1.
\end{enumerate}

\noindent This procedure will control the FWER under any joint distribution of the test statistics. 

\end{document}